\documentclass[12pt,letterpaper]{article}
\usepackage[bf]{caption}

\usepackage{morefloats}

\usepackage[margin=1.5in, right=1.5in, top=1.5in, bottom=1.5in]{geometry}
\usepackage{layout}
\usepackage{natbib}
\bibpunct{(}{)}{;}{a}{}{,}
\usepackage{array}
\usepackage{fontenc}
\usepackage{longtable}
\usepackage{graphicx}
\usepackage{epstopdf}
\usepackage{amsmath}
\usepackage{amssymb}
\usepackage{amsthm}
\usepackage{color}
\usepackage{float}
\usepackage{epsfig}
\usepackage{epsf}
\usepackage{color}
\usepackage{rotating}
\usepackage{lscape}
\usepackage{longtable}
\usepackage{latexsym}
\usepackage[rightcaption]{sidecap}
\usepackage{changes}
\colorlet{Changes@Color}{red}

\usepackage{times}
\fontfamily{ptm}\selectfont

\makeatletter
\renewcommand\section{\@startsection{section}{1}{\z@}%
																		{-3.5ex \@plus -1ex \@minus -.2ex}%
																		{2.3ex \@plus.2ex}%
																		{\normalfont\large\bfseries}}
\makeatother
\newcommand{\at}{\makeatletter @\makeatother}

  \setkeys{Gin}{draft=false}

\begin{document}

\begin{center}
Evolution of Earth-like  planetary atmospheres around M-dwarf stars:\\
Assessing the atmospheres and biospheres with a
     coupled atmosphere biogeochemical model
\end{center}

\vspace*{3mm}

\begin{center}
 S. Gebauer$^{1,2}$, J. L. Grenfell$^2$, R. Lehmann$^3$, and H. Rauer$^{1,2}$
\end{center}

\vspace*{3mm}

\begin{center}
(1) Zentrum f\"ur Astronomie und Astrophysik (ZAA), Technische Universit\"at Berlin (TUB), Hardenbergstr. 36, 10623 Berlin, Germany\\
\vspace*{1mm}
(2) Institut f\"ur Planetenforschung (PF), Abteilung Extrasolare Planeten und Atmosph\"aren (EPA), Deutsches Zentrum f\"ur Luft- und Raumfahrt (DLR), Rutherfordstr. 2, 12489 Berlin, Germany\\

(3) Alfred-Wegener Institut, Helmholtz-Zentrum f\"ur Polar- und Meeresforschung, Telegrafenberg A45, 14473 Potsdam, Germany\\

\end{center}

\vspace*{10mm}

\begin{center}
Running title: Earth-like atmosphere evolution around M-dwarf stars
\end{center}

\newpage
\begin{center}
Corresponding author:\\ Dr. Stefanie Gebauer\\ Institut f\"ur Planetenforschung (PF)\\ Abteilung e Planeten und Atmosph\"aren (EPA)\\
Deutsches Zentrum f\"ur Luft- und Raumfahrt (DLR)\\ Rutherfordstr. 2\\ 12489 Berlin\\ Germany\\ phone: +49-(0)30-67055-454\\ e-mail: stefanie.gebauer\at dlr.de
\end{center}

\vspace*{10mm}

\noindent "Peer-review version of the manuscript. Final publication is available from Mary Ann Liebert, Inc., publishers http://doi.org/10.1089/ast.2017.1723"

\newpage

\begin{abstract}
\normalsize 
Earth-like planets orbiting M-dwarf stars are prominent future targets when searching for life outside the solar system.
We apply our newly developed Coupled Atmosphere Biogeochemistry (CAB) model to investigate the coupling between the biosphere, geosphere and atmosphere in order to gain deeper insight into the atmospheric evolution of Earth-like  planets orbiting M-dwarf stars. Our main goal is to understand better atmospheric processes affecting biosignatures and climate on such worlds.
Furthermore, this is the first study to our knowledge which applies an automated chemical pathway analysis quantifying the production and destruction pathways of molecular oxygen (O$_2$) for an Earth-like  planet with an Archean O$_2$ abundance orbiting in the habitable zone of the M-dwarf star AD Leonis (AD Leo), which we take as a type-case of an active M-dwarf star.
Results suggest that the main production arises in the upper atmosphere from carbon dioxide (CO$_2$) photolysis followed by catalytic hydrogen oxide radical (HO$_x$) reactions. The strongest destruction does not take place in the troposphere, as was the case in \citet{gebauer2017} for an early-Earth analog planet around the Sun, but instead in the middle atmosphere where water (H$_2$O) photolysis is the strongest. This result was driven by the strong Lyman-$\alpha$-radiation output of AD Leo, which efficiently photolyzes H$_2$O.
Results further suggest that early Earth-like atmospheres of planets orbiting an M-dwarf star like AD Leo are in absolute terms less destructive for atmospheric O$_2$ than for early-Earth analog planets around the Sun despite higher concentrations of reduced gases such as e.g. H$_2$, CH$_4$ and CO. 
Hence the net primary productivity required to produce the same amount of atmospheric O$_2$ at the surface is reduced. This result was due to a change in the atmospheric oxidative capacity, driven by the input stellar spectrum resulting in shifts in the intra-familiy HO$_x$ partitioning. This implies that a possible "Great Oxidation event", analogous to that on Earth, would have occured earlier in time in analog atmospheres around M-dwarf stars, assuming the same atmospheric O$_2$ sources and sinks as on the Early Earth. For an Earth-like planet around an active M-dwarf, smaller amounts of net primary productivity (NPP) could therefore be required to oxygenate the atmosphere compared to Earth around the Sun.

\end{abstract}

\newpage

\section{Introduction}
The study of \citet{gebauer2017} (hereafter paper I) addressed the evolution of early Earth analog planets around the Sun and the implications for their atmospheres and biospheres by applying a Coupled Atmosphere Biogeochemical (CAB) model. In the present paper we investigate Earth-like  planets with differing contents of molecular oxygen (O$_2$) orbiting in the Habitable Zone (HZ) of the M-dwarf star AD Leonis (AD Leo), which we take as a type-case for an active M-dwarf host star. 

M-dwarf stars make up about 75\% of all stars in the galaxy \citep{tarter2007} and are favored observational targets. Due to their very low masses, the time which they spend on the main sequence theoretically exceeds the lifetime of the universe \citep{reid2005}. Their HZ lies close-in due to the low stellar luminosity. An increasing number of so-called super-Earths have been detected orbiting such stars (e.g. \citealt{gillon2016,affer2016,gillon2017}). The newly discovered planet with 1.3 Earth masses orbiting Proxima centauri \citep {anglada2016} is a promising candidate for an Earth-like  planet in the HZ of an M-dwarf star.

Due to the stellar characteristics of M-dwarf stars however, several issues arise related to the potential habitability of planets in their HZs.
These stars show a strong temporal variability and are usually very active emitting large amounts of (E)UV radiation and X-rays (see \citealt{segura2005}). Coronal mass ejections are a frequent phenomenon for active M-dwarf stars which may erode the planetary atmosphere \citep{khoda2007,lammer2007,lammer2008}. M-dwarf stars exhibit a dense and fast stellar wind \citep{vidotto2015}. Due to the very close-in HZ planets are likely to become tidally-locked within a short time after formation \citep{dole1964,kasting1993,selsis2007} leading to a possible atmospheric collapse \citep{haberle1996,joshi1997,joshi2003,word2015} on the night side or evaporation on the day-side. It was shown by \citet{yang2013,yang2014} that the inner HZ edge could be extended towards the star due to synchronous rotation of the planet coupled with cloud formation. However, this result was contested by a recent study of \citet{kopp2016} who applied a general circulation model with self-consistent relationships between stellar metallicity, stellar effective temperature, and the planetary orbital/rotational period.  

Earth-like planets around M-dwarf stars might exhibit possibly weak planetary magnetic fields, hence experience potentially strong fluxes of cosmic rays which could have an impact on the atmosphere \citep{lammer2007,khoda2007} and on biosignature signals \citep{griess2010,grenfell2012cr,taba2016}. More recent 3D model studies of Earth-like planets orbiting M-dwarfs focus on determining habitability see e.g. \citet{boutle2017,wolf2017} and references therein. 

M-dwarf star planets have a deficiency in Photosynthetically Active Radiation (PAR) (which lies predominantly in the wavelength range of 400 - 720 nm) because PAR emitted by such stars is weakened by e.g. photospheric absorption of TiO. The solar flux at the surface of the Earth has a maximum at 685 nm whereas for AD Leo the peak surface flux assuming Earths atmosphere lies around 1045 nm resulting in 0.125 times the Earth's 400 - 700 nm PAR band\footnote{For Earth the global, time-averaged surface incident photon flux is $2.4\cdot10^{20}$ photons/m$^2$/s.} \citep{kiang2007}. That study showed that photosynthesis could also proceed at wavelengths longer than 720 nm. 

For comprehensive reviews of occurrence and habitability of planets orbiting M-dwarf stars see \citet{scalo2007} and \citet{shields2016}.

\begin{figure}
\centering
\includegraphics[width=8.5cm]{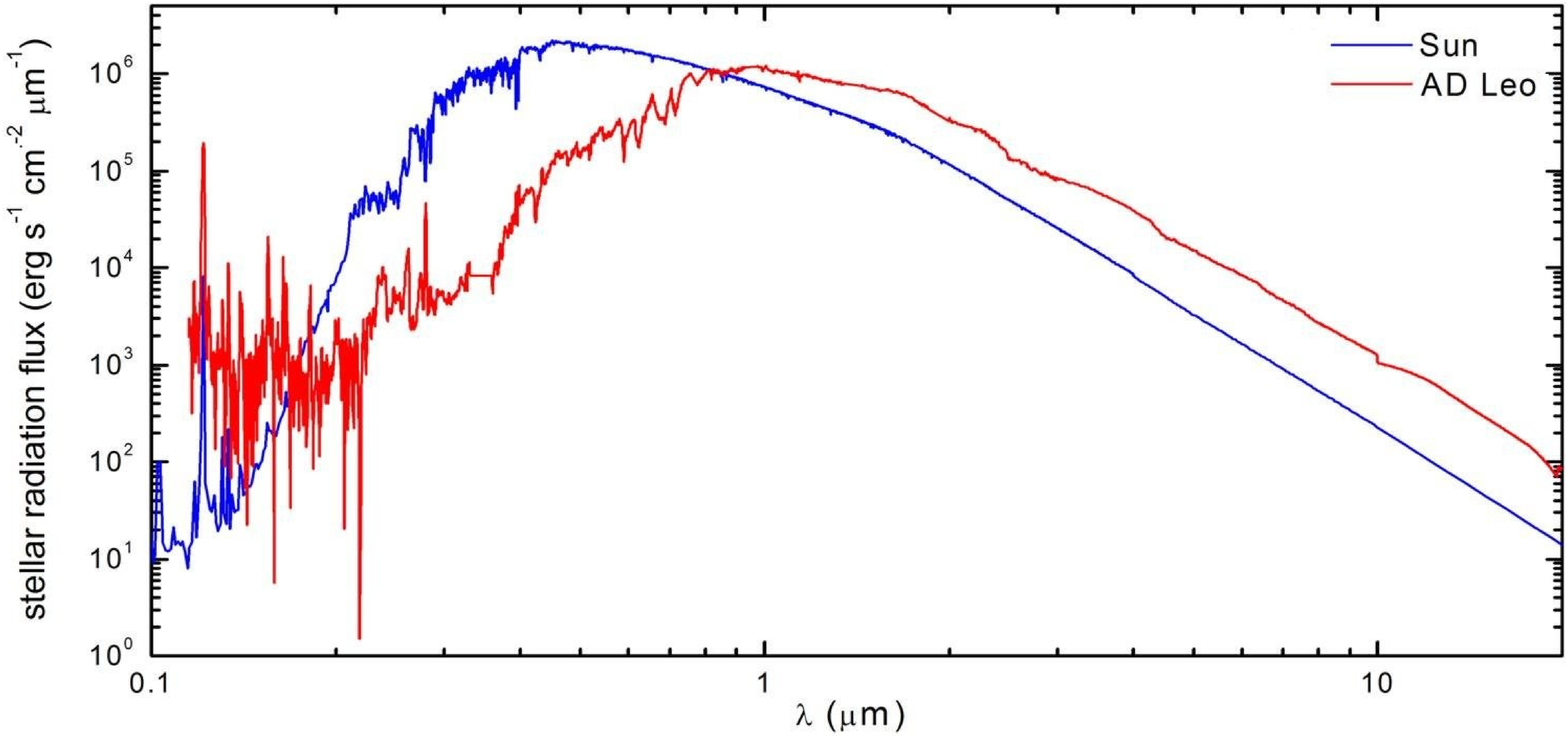}
  \caption{Stellar radiation flux of the M-dwarf star AD Leo in red (derived from IUE, International Ultraviolet Explorer, \citealp{pettersen1989,Leggett1996} and \citealp{hauschildt1999}) in comparison to the Sun in blue \citep{Gueymard2004}.}
  \label{adlspecfig}
\end{figure}

In this paper we focus on Earth-like  planets orbiting in the habitable zone at 0.153 AU (Astronomical Unit) around the well-studied M-dwarf star AD Leo where the total stellar wavelength-integrated energy input equals that of 1 Solar Constant (1366 W m$^{-2}$). Fig. \ref{adlspecfig} shows the stellar spectrum of AD Leo in comparison to the Sun illustrating that the stellar flux of AD Leo is generally higher in the IR whereas in the visible it is reduced in comparison to the Sun. Furthermore, AD Leo shows high stellar activity in the (E)UV range.



\subsection{Contribution of this Work}
In this work we will investigate different evolutionary stages of hypothetical Earth-like planets orbiting AD Leo using Earth as a reference system and applying the CAB model. Ours is the first study to our knowledge which considers detailed coupled climate-photochemistry, including interactive atmospheric O$_2$, CO$_2$ and N$_2$ and including biogeochemical processes for Earth-like atmospheres around M-dwarf stars. 
Assuming the same biological and geological evolution as on Earth we will calculate the impact of the different stellar environment on the NPP of oxygenic photosynthesis. 

For the first time we investigate atmospheric chemical responses of O$_2$ in an Earth-like  planetary atmosphere for an Archean type atmosphere (with $10^{-6}$ PAL O$_2$) by applying the Pathways Analysis Program (PAP) developed by \citet{lehmann2004}.

The paper is structured as follows:
Section \ref{cabmodelsection} and Section \ref{papsection} describe briefly the models used. Section \ref{scenariodescription} gives an overview of the simulated Earth-like  planet scenarios around AD Leo whereas in Section \ref{resultssection} the results are presented and discussed. Our main conclusions are presented in Section \ref{resultsection}.

\section{Coupled Atmosphere Biogeochemical (CAB) Model}\label{cabmodelsection}
In order to address the interactions between atmospheric, geological, chemical and biotic systems the Coupled Atmosphere Biogeochemical (CAB) model as discussed in paper I is applied. The CAB model is composed of two submodules: 1. atmosphere and 2. biogeochemistry (based on \citealp{goldblatt2006}). Both submodules are coupled via the stagnant boundary layer model \citep{liss1974,kharecha2005}. 
The 1D global mean cloud-free, steady state atmospheric module based on e.g. \citet{kasting1984,segura2003,grenfell2007cr,grenfell2007,rauer2011,grenfell2011} is composed of a climate and a photochemistry part calculating temperature, water and chemical species profiles. Unlike earlier works, the concentrations of O$_2$, CO$_2$ and N$_2$ are not simply held constant to fixed isoprofile values but are instead calculated interactively in the chemistry. All relevant atmospheric sources and sinks of O$_2$ are implemented, e.g. volcanic and metamorphic outgassing of reduced chemical species according to geological time, $t_{\rm{geo}}$, and hydrogen escape at the top-of-the-atmosphere (TOA). For a more detailed description we refer to paper I.

\section{Pathways Analysis Program}\label{papsection}
The Pathway Analysis Program (PAP) was developed by \citet{lehmann2004} in order to automatically identify chemical pathways in any arbitrary chemical reaction network and to quantify their efficiencies by assigning rates. PAP was applied by e.g. \citet{grenfell2006,verronen2011,stock2012a,stock2012b,grenfell2013,verronen2013,stock2017} and in paper I. The algorithm yields a list of all dominant pathways which produce, destroy or recycle a chemical species of interest. For a more detailed description of PAP see paper I and \citet{lehmann2004}.
%

For the analysis of large and complex reaction networks chemical pathways with a rate below a user-defined threshold rate $f_{\rm{min}}$ are deleted in order to avoid long computational time. In the present study, $f_{\rm{min}}=10^{-12}$ parts per billion by volume per second (ppbv/s)) was chosen which is sufficient for finding the dominant pathways producing or consuming atmospheric O$_2$ in our simulated atmosphere. A PAP analysis was performed for each of the 64 vertical layers of the atmospheric column module. The resulting production and destruction rates of O$_2$ from each individual pathway are integrated over the atmospheric module vertical grid and are expressed as a percentage of the total column-integrated production and destruction rate from all pathways found by PAP.


\section{Application to Early Earth Analog Atmospheres around AD Leo}\label{scenariodescription}
For simulating  planets at different evolutionary stages the O$_2$ volume mixing ratio of the atmosphere at the surface is taken from the evolutionary path of O$_2$ adapted from \citet{catling2005} (shown in paper I, their Fig. 3). We do not assume any luminosity change of AD Leo because this stays relatively constant over its evolution \citep{reid2005}. 

For all runs we assume:
\begin{itemize}
	\item AD Leo input spectrum,
	\item surface pressure $p_{\rm{surf}}=1$ bar with N$_2$ as a fill-gas,
	\item gravity acceleration $g=981$ cm s$^{-2}$,
	\item biogenic surface fluxes of CH$_4$ (474 Tg/yr), CO (1796 Tg/yr), N$_2$O (13.5 Tg/yr), CH$_3$Cl (3.4 Tg/yr) (these values lead to modern Earth abundances for the given species), deposition velocities of all other chemical compounds as in paper I,
	\item crustal mineral redox buffer is set to quartz-fayalite-magnetite (QFM) ($\Delta f_{\rm{O}_2}=0$),
	\item surface vmr of CO$_2$ of 355 ppm.
	\end{itemize}

Volcanic emissions of H$_2$, H$_2$S, SO$_2$, CO, CH$_4$, and CO$_2$ and metamorphic emissions of H$_2$ and CH$_4$ are varied according to the geological time $t_{\rm{geo}}$.


\section{Results and Discussion}\label{resultssection}
\subsection{Atmosphere Modeling}
\subsubsection{Climate Responses}
\begin{figure}
\centering
\includegraphics[width=8.5cm]{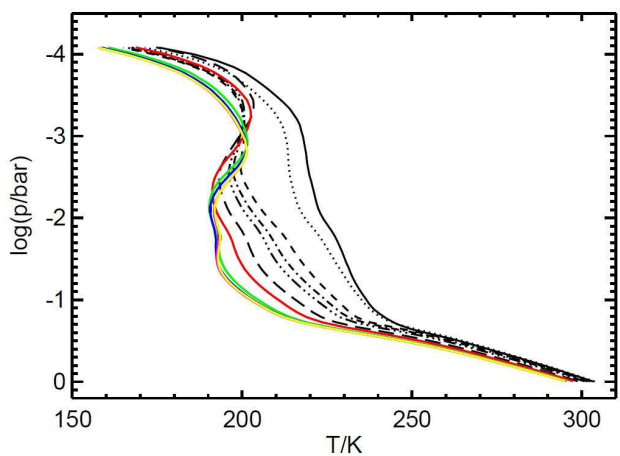} 
  \caption{Selected p-T profiles of Earth-like  atmosphere runs around the M-dwarf star AD calculated by the CAB model. Notation: solid black: 1 PAL O$_2$ (modern Earth), dotted: $2.5\cdot 10^{-1}$ PAL ($t_{\rm{geo}}=0.82$ Gyrs), short dashed: $10^{-1}$ PAL ($t_{\rm{geo}}=2.18$ Gyrs), dash-dot: $2.5\cdot 10^{-2}$ PAL ($t_{\rm{geo}}=2.22$ Gyrs), dash-dot-dot-dot: $10^{-2}$ PAL ($t_{\rm{geo}}=2.22$ Gyrs), long dashed: $2.5\cdot 10^{-3}$ PAL ($t_{\rm{geo}}=2.23$ Gyrs), red: $10^{-3}$ PAL ($t_{\rm{geo}}=2.24$ Gyrs), green: $2.5\cdot 10^{-4}$ PAL ($t_{\rm{geo}}=2.25$ Gyrs), blue: $10^{-4}$ PAL ($t_{\rm{geo}}=2.28$ Gyrs), magenta: $10^{-5}$ PAL ($t_{\rm{geo}}=2.59$ Gyrs), yellow: $10^{-6}$ PAL ($t_{\rm{geo}}=2.69$ Gyrs).}
  \label{figure_pTvarO2}
\end{figure}

The $p$-$T$ profiles for selected Earth-like  atmosphere runs around AD Leo are shown in Fig. \ref{figure_pTvarO2}. 
For the 1 PAL O$_2$ atmospheric run stratospheric temperatures are reduced
in comparison to modern Earth around the Sun (see Fig. 4 in Paper I) mainly due to the reduced UV flux longwards of about 200 nm (see Fig. \ref{adlspecfig}) which results in less radiative heating by ozone (O$_3$). On reducing the atmospheric O$_2$ content, i.e. going backwards in time, the stratospheric temperatures shown in Fig. \ref{figure_pTvarO2} decrease
by about 40 K due to overall less atmospheric heating by O$_3$.
A weak temperature inversion develops at about 0.01 bar due to increased heating
by O$_3$ and less cooling by CO$_2$ in the upper stratosphere of Earth-like  atmosphere runs with lower O$_2$ contents as a result of enhanced O$_3$ concentrations (see Fig. \ref{figure_O3varO2}) and decreasing CO$_2$ mixing ratios towards lower atmospheric pressures (see Fig. \ref{CO2varO2fig}).

\begin{figure}
\centering
\includegraphics[width=8.5cm]{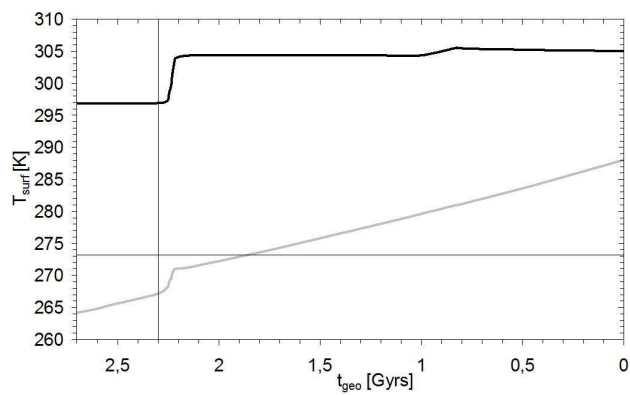} 
  \caption{Surface temperature for calculated Earth-like  atmosphere runs around the M-dwarf star AD Leo (shown in solid black) in comparison to the early-Earth analog planets around the Sun presented in paper I (shown in solid grey). The same evolution of surface O$_2$ concentration is assumed. Additionally indicated as a horizontal line is the freezing temperature of H$_2$O of $T = 273.15$ K. The beginning of the GOE at $t_{\rm{geo}} = 2.3$ Gyrs is indicated as a vertical line.}
  \label{figure_Tdecrease}
\end{figure}

On going forwards in time, i.e. increasing O$_2$ surface concentrations, surface temperatures (see Fig. \ref{figure_Tdecrease}) increase from 296.9 K before the GOE at 10$^{-6}$ PAL O$_2$ to 305.1 K at 1 PAL O$_2$. Generally, surface temperatures are higher than for runs around the Sun firstly, due to less Rayleigh scattering and secondly, due to an enhanced greenhouse effect from molecular absorption by CH$_4$ and H$_2$O which are both enhanced in these atmospheres (see Section \ref{photochem}). Since total stellar luminosity is held constant we do not find an overall increase of surface temperature with time as observed for the early-Earth analog planets around the Sun. Furthermore, we observe a small increase in surface temperature around the Second Oxidation Event (SOE). Further analysis suggested this was related to a decrease in hydroxyl (OH) which has a complex photochemistry. This in turn led to an increase in the greenhouse gas methane (CH$_4$) hence surface warming and a rather strong increase in H$_2$O via evaporation around that time (see Section \ref{photochem}).
For the low O$_2$ atmosphere run around AD Leo shortwave radiation is less absorbed, whereas Rayleigh scattering is increased and greenhouse warming is reduced compared to the 1 PAL O$_2$ run. 

As in paper I we will now validate the net integrated TOA fluxes calculated by our long-wave radiation module RRTM \citep{mlawer1997} against the line-by-line (lbl) radiative transfer code SQuIRRL (Schwarzschild Quadrature InfraRed Radiation Line-by-line; developed by \citealp{schreier2001,schreier2003,schreier2014}). For all runs considered a maximum relative change of 1.0 \% was found.


\subsubsection{Photochemical Responses}\label{photochem}
\begin{figure}
\centering
\includegraphics[width=8.5cm]{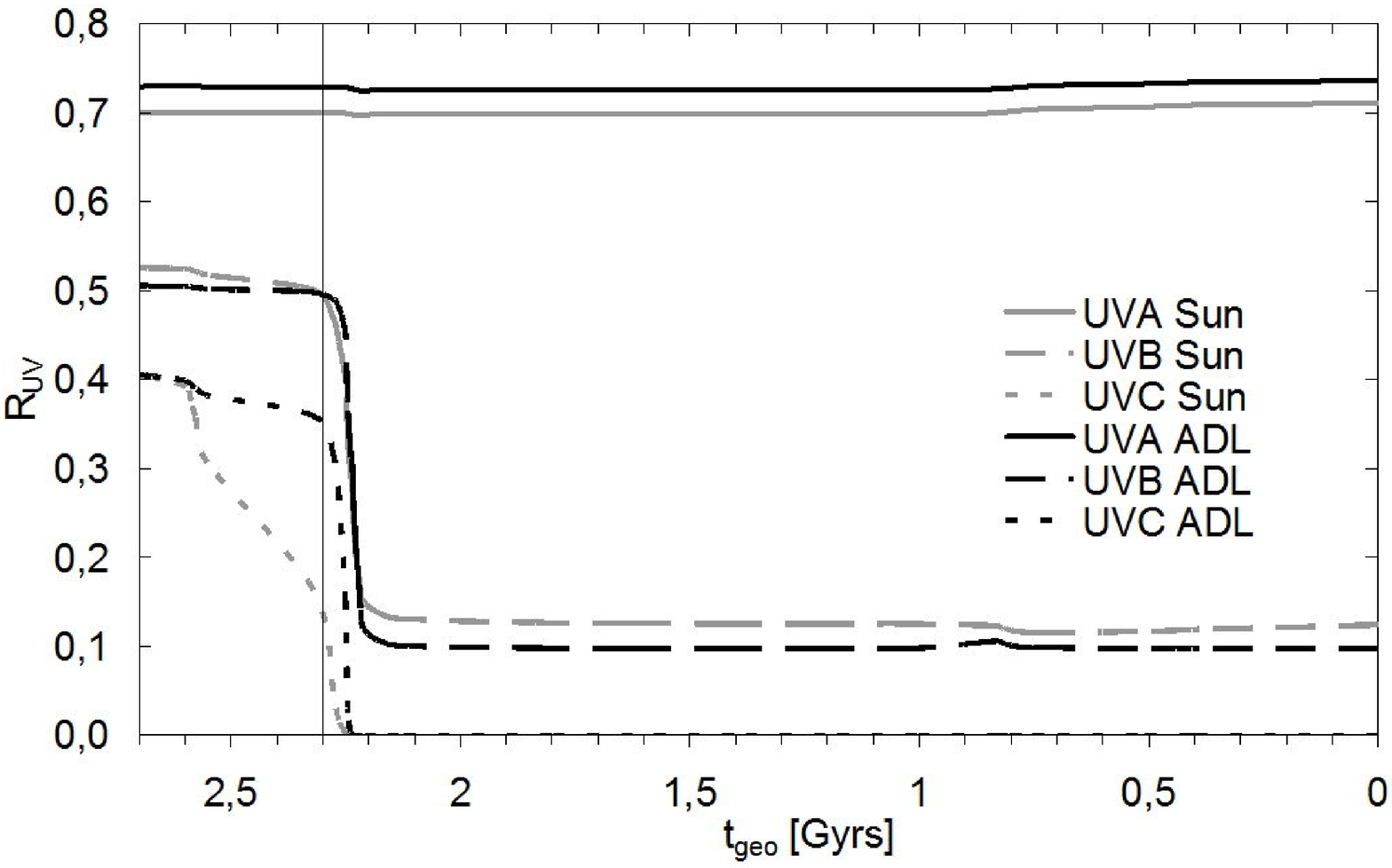} 
  \caption{Ratio (surface/TOA) radiation fluxes ($R_{UV}$) for UVA (solid), UVB (long dashed) and UVC (short dashed) for Earth-like extrasolar atmosphere runs around the M-dwarf star AD Leo (shown in black) as a function of geological time $t_{\rm{geo}}$ in comparison to the early-Earth analog planets around the Sun presented in paper I (shown in grey). The beginning of the GOE at $t_{\rm{geo}} = 2.3$ Gyrs is indicated as a vertical line.}
  \label{figure_ruv}
  \end{figure}

\paragraph{UV Radiation environment}
Generally, for an Earth-like  planet orbiting in the HZ of the M-dwarf AD Leo the UV radiation environment is less strong than compared to Earth around the Sun. For the AD Leo scenarios at the surface, UVA radiation is virtually constant on going forwards in time. One calculates before the GOE 1.49 W m$^{-2}$ and afterwards 1.48 W m$^{-2}$. UVB radiation is enhanced at the surface before the GOE with 0.16 W m$^{-2}$ in comparison to atmospheres after the GOE with 0.02 W m$^{-2}$. For the UVC radiation\footnote{Here, UVC radiation denotes the wavelength range from 253-280 nm.} at the surface before the GOE we calculate 0.13 W m$^{-2}$ whereas afterwards it is virtually zero. 
Fig. \ref{figure_ruv} shows the radiation shielding efficiency, $R_{UV}$, (i.e. the ratio (surface/TOA) radiation) of such atmospheres in comparison with an early-Earth analog planet around the Sun. Low $R_{UV}$ values denote high shielding efficiency. UVA and UVC radiation is shielded less efficiently (higher $R_{UV}$) for the AD Leo atmosphere scenarios whereas for UVB, the shielding efficiency is higher.

\begin{figure}
\centering
\includegraphics[width=8.5cm]{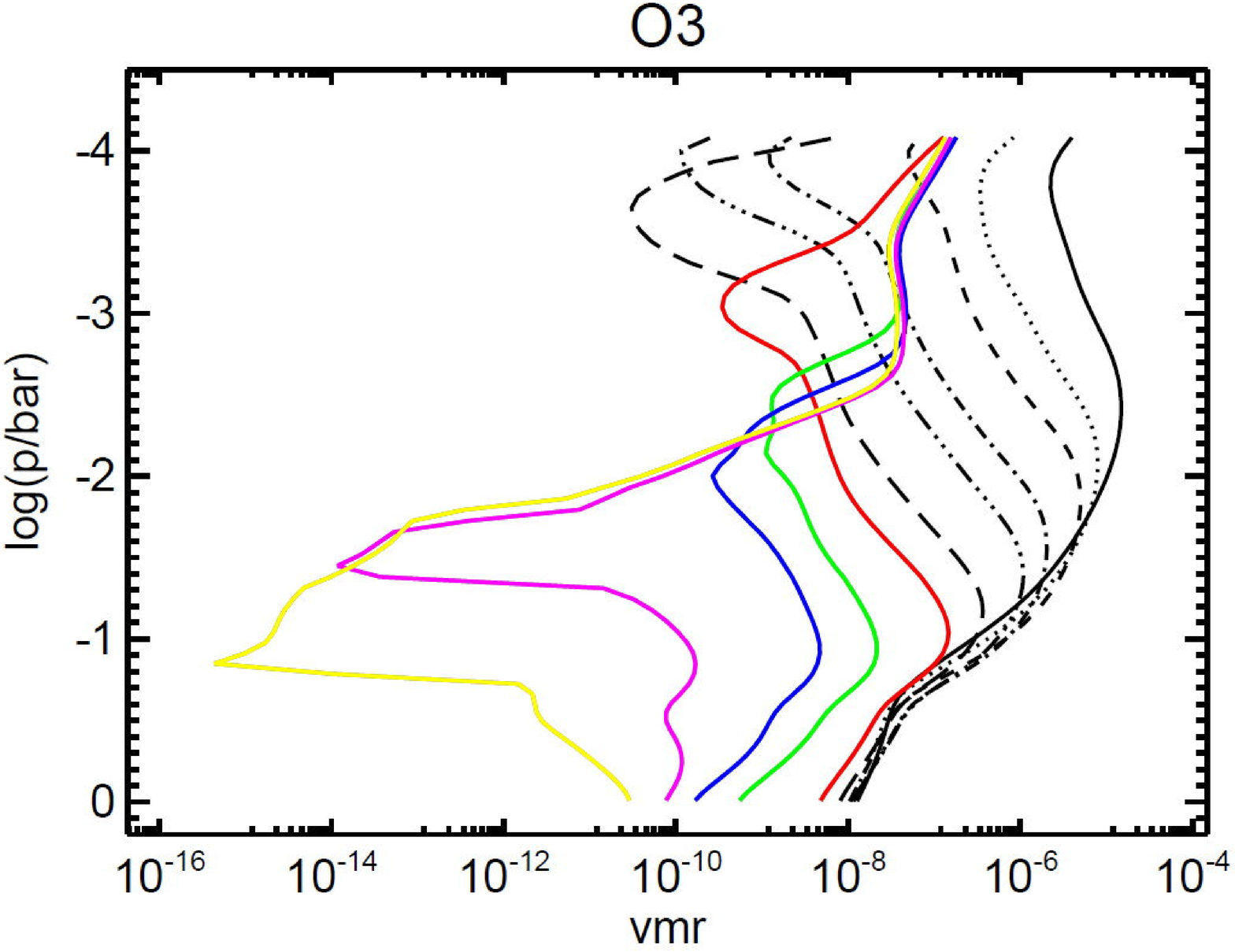} 
  \caption{O$_3$ profiles of Earth-like  atmosphere runs around the M-dwarf AD Leo calculated by the CAB model. Notation as in Fig. \ref{figure_pTvarO2}.}
  \label{figure_O3varO2}
\end{figure}

\begin{figure}
\centering
\includegraphics[width=8.5cm]{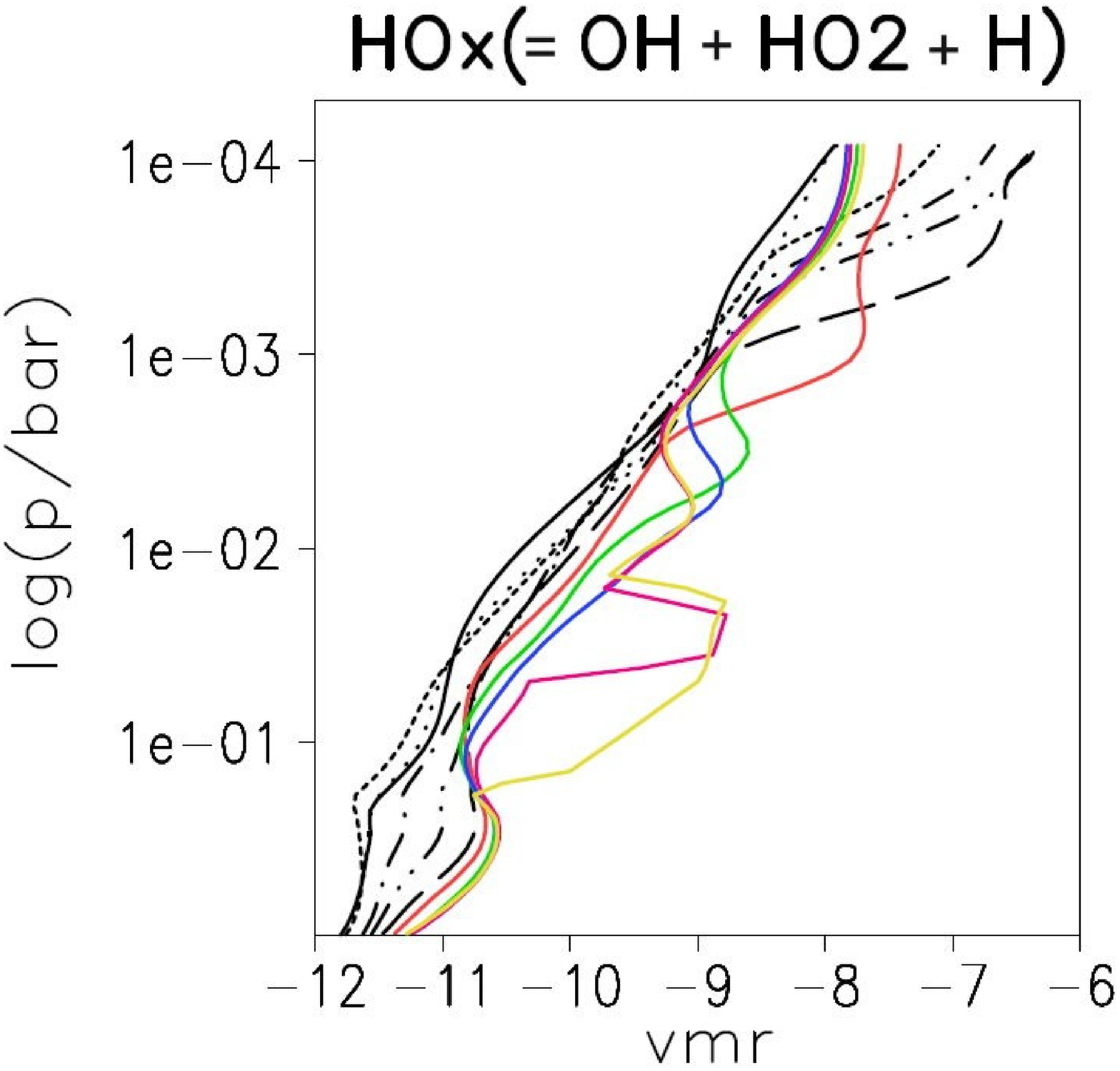}
  \caption{HO$_x$ (= OH + HO$_2$ + H) profiles of Earth-like  atmosphere runs around the M-dwarf AD Leo calculated by the CAB model. Notation as in Fig. \ref{figure_pTvarO2}.}
  \label{HOxADL2fig}
\end{figure}

\begin{figure}  
 \centering
\includegraphics[width=8.5cm]{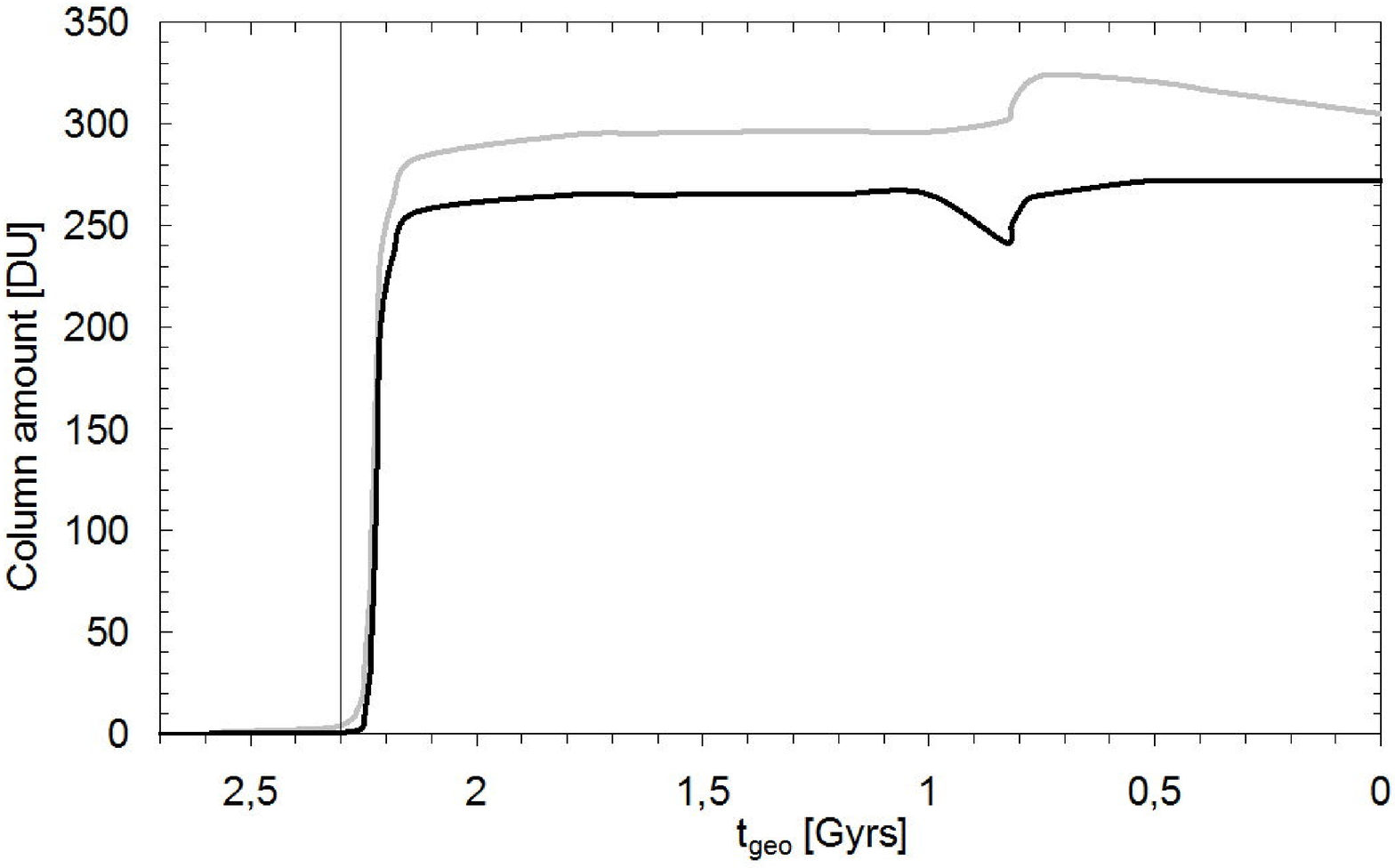} 
  \caption{O$_3$ column amount [DU] for Earth-like  atmosphere runs around the M-dwarf star AD Leo (black) as a function of geological time $t_{\rm{geo}}$ in comparison to the early-Earth analog planets around the Sun presented in paper I (shown in grey). The beginning of the GOE at $t_{\rm{geo}} = 2.3$ Gyrs is indicated as a vertical line.}
  \label{figure_O3column}  
\end{figure}

In the following text we focus on the photochemical analysis of important atmospheric constituents such as proposed biosignatures (O$_2$, O$_3$, N$_2$O) and related compounds (CH$_4$, H$_2$O, CO$_2$) as in paper I. 
\paragraph{\textbf{Ozone - O$_3$}}
Fig. \ref{figure_O3varO2} presents selected O$_3$ profiles of Earth-like  atmosphere runs around the M-dwarf AD Leo.
For an Earth-like  planet around AD Leo with 1 PAL O$_2$ the O$_3$ profile is comparable to that on modern Earth. Similarly to scenarios presented in paper I, the O$_3$ maximum moves to higher pressures on reducing the O$_2$ surface mixing ratio due to the ozone-UV feedback as discussed in paper I. For atmospheres with ground level O$_2$ concentrations at and below 10$^{-3}$ PAL (i.e. the solid colored lines in Fig. \ref{figure_O3varO2}) stratospheric O$_3$ starts to strongly increase towards lower pressures because of enhanced O$_2$ concentrations due to CO$_2$ photolysis (see Fig. \ref{O2varO2fig} and Section \ref{papanalysis} for chemical analysis). Below $2.5\cdot 10^{-5}$ PAL O$_2$ a local O$_3$ minimum becomes more pronounced and penetrates deeper into the atmosphere as O$_2$ is reduced. This is related firstly, to reduced O$_2$ abundances (see again Fig. \ref{O2varO2fig}) and secondly, due to enhanced HO$_{\mbox{x}}$ concentrations (= OH + HO$_2$ + H, shown in Fig. \ref{HOxADL2fig}) which catalyze O$_3$ loss in the middle atmosphere. Thereby, H is the dominant species which is increasingly produced by H$_2$O photolysis in the pressure regime between 0.1 and 0.01 bar via H$_2$O + h$\nu$ $\rightarrow$ H + OH (see below). Additional photolytic destruction of CO$_2$ via CO$_2$ + h$\nu$ $\rightarrow$ CO + O and subsequent reaction between OH and CO via CO + OH $\rightarrow$ H + CO$_2$ (not shown) represents an additional source for H.
Fig. \ref{figure_O3column} suggests that in Earth-like atmospheres around the M-dwarf AD Leo compared to early-Earth analog planets around the Sun the O$_3$ column amount is generally reduced after the GOE. A small minimum arises around the SOE due to an increase in surface temperature, hence H$_2$O and therefore HO$_{\mbox{x}}$. Normally, O$_3$ is considered to be a good proxy for O$_2$ (see e.g. \citep{segura2003}). Fig. \ref{figure_O3column} however, suggests that this assumption may not always be valid since around the SOE O$_3$ decreases with increasing O$_2$ for the AD Leo scenarios. This further illustrates the sensitivity of OH photochemistry.

\begin{figure}
\centering
\includegraphics[width=8.5cm]{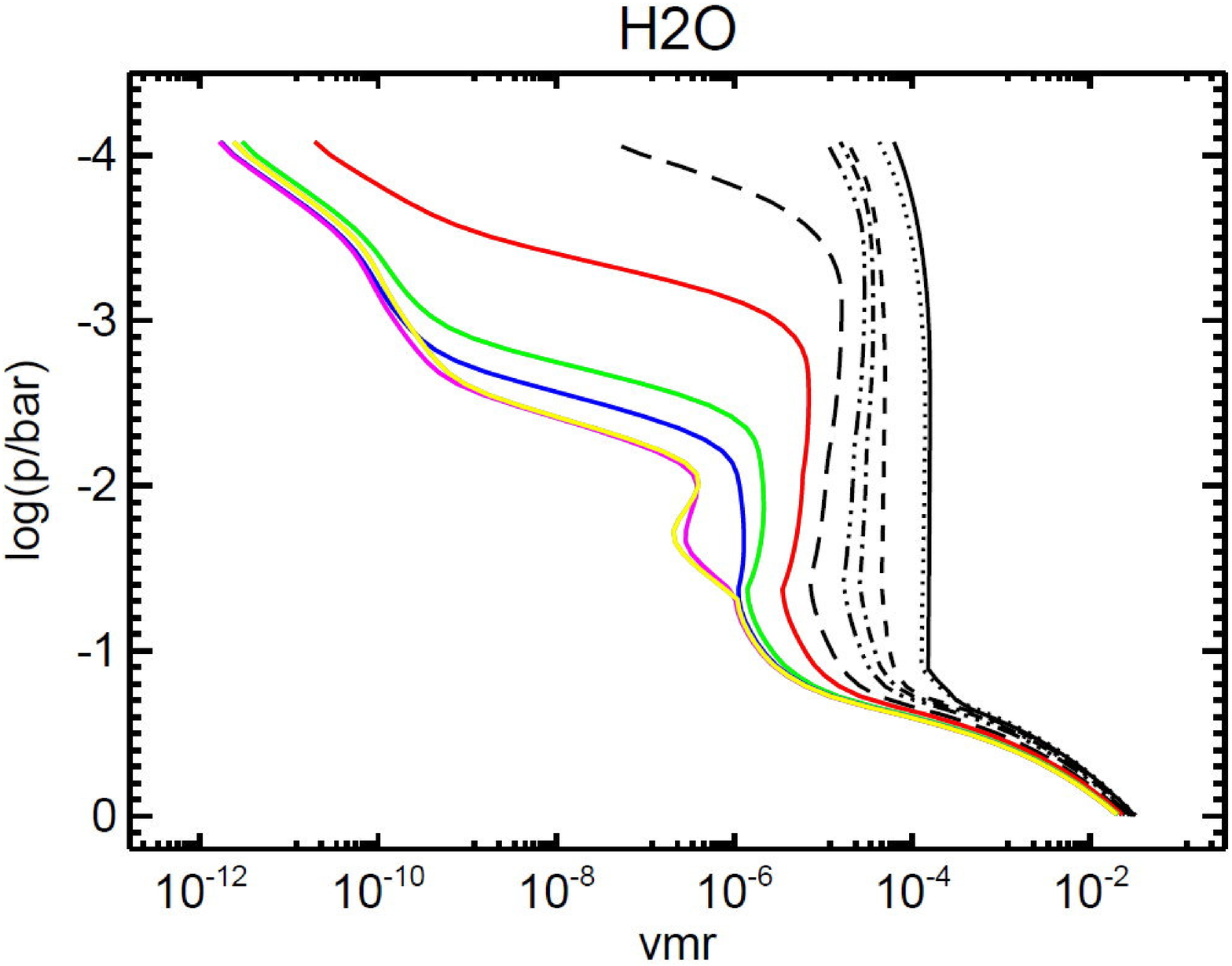} 
  \caption{H$_2$O profiles of Earth-like  atmosphere runs around the M-dwarf AD Leo calculated by the CAB model. Notation as in Fig. \ref{figure_pTvarO2}.}
  \label{H2OvarO2fig}
\end{figure}
\begin{figure}
\centering
\includegraphics[width=8.5cm]{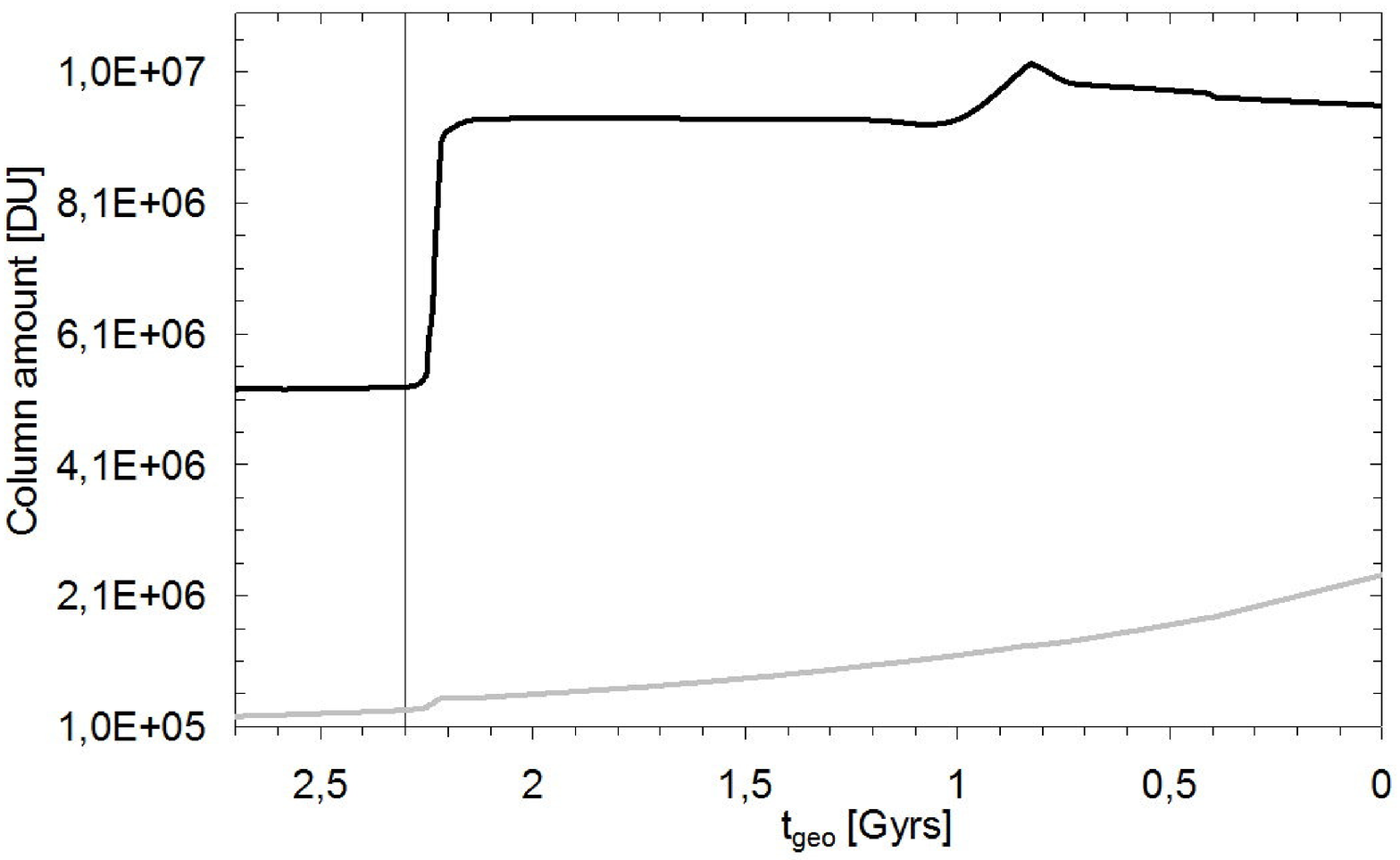} 
  \caption{H$_2$O column amount [DU] for Earth-like  atmosphere runs around the M-dwarf star AD Leo (black) as a function of geological time $t_{\rm{geo}}$ in comparison to the early-Earth analog planets around the Sun presented in paper I (shown in grey). The beginning of the GOE at $t_{\rm{geo}} = 2.3$ Gyrs is indicated as a vertical line.}
  \label{H2Ocolumnfig} 
\end{figure}

\paragraph{\textbf{Water - H$_2$O}}
H$_2$O profiles for the same atmospheric scenarios are presented in Fig. \ref{H2OvarO2fig}. In comparison to the scenarios presented in paper I around the Sun, the tropospheres of Earth-like  atmosphere scenarios around AD Leo are generally more wet due to more evaporation of H$_2$O via a large CH$_4$ greenhouse (see below). For O$_2$ concentrations above 0.01 PAL O$_2$ the H$_2$O mixing ratio stays approximately constant with height above the tropopause. Below 0.01 PAL O$_2$ H$_2$O mixing ratios decrease in the upper stratosphere because of strong photolytic destruction. The overall column amount of H$_2$O, shown in Fig. \ref{H2Ocolumnfig}, is increased by up to an order of magnitude in comparison to the scenarios presented in paper I. After an increase around the GOE the H$_2$O column amounts stay relatively constant after the GOE with a second increase around the SOE.

\begin{figure}
\centering
\includegraphics[width=8.5cm]{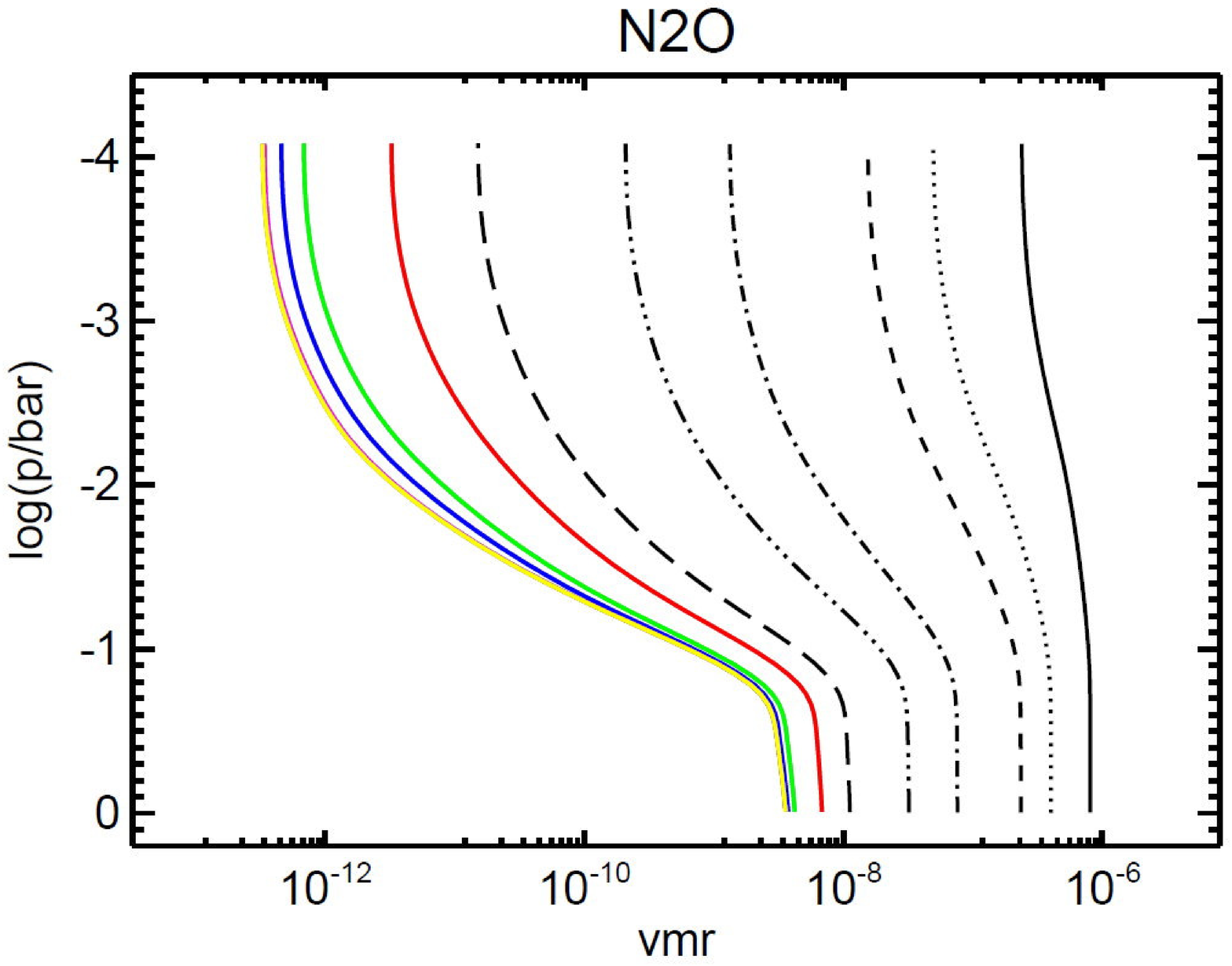} 
  \caption{N$_2$O profiles of Earth-like  atmosphere runs around the M-dwarf AD Leo calculated by the CAB model. Notation as in Fig. \ref{figure_pTvarO2}.}
  \label{N2OvarO2fig}
\end{figure}
\begin{figure}  
\centering
\includegraphics[width=8.5cm]{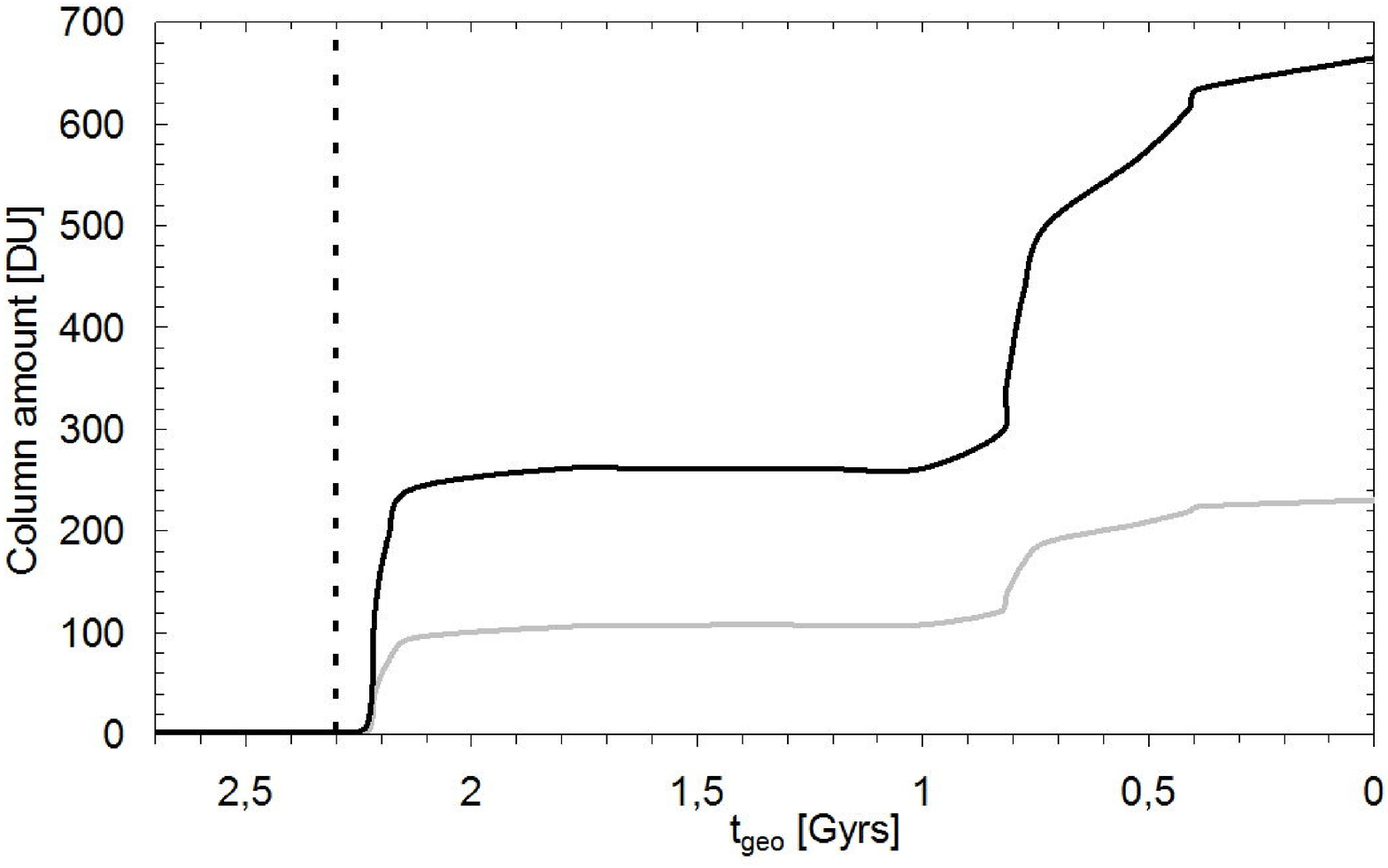} 
  \caption{N$_2$O column amount [DU] for Earth-like  atmosphere runs around the M-dwarf star AD Leo (black) as a function of geological time $t_{\rm{geo}}$ in comparison to the early-Earth analog planets around the Sun presented in paper I (shown in grey). The beginning of the GOE at $t_{\rm{geo}} = 2.3$ Gyrs is indicated as a vertical line.}
  \label{N2Ocolumnfig}
\end{figure}

\paragraph{\textbf{Nitrous oxide - N$_2$O}}
On going forwards in time N$_2$O mixing ratios steadily increase as shown in Fig. \ref{N2OvarO2fig}. Compared to the scenarios in paper I, N$_2$O concentrations in atmospheres of Earth-like  atmosphere scenarios around AD Leo are enhanced which is also reflected in the total column amount shown in Fig. \ref{N2Ocolumnfig}. This arises due to much weaker UVB emissions from AD Leo compared with the Sun (see Fig. \ref{adlspecfig}) and weaker excited oxygen [O($^1$D)] which comes from weaker O$_3$ photolysis (see below) since both UVB and  O($^1$D) are major sinks for N$_2$O.

\begin{figure}
\centering
\includegraphics[width=8.5cm]{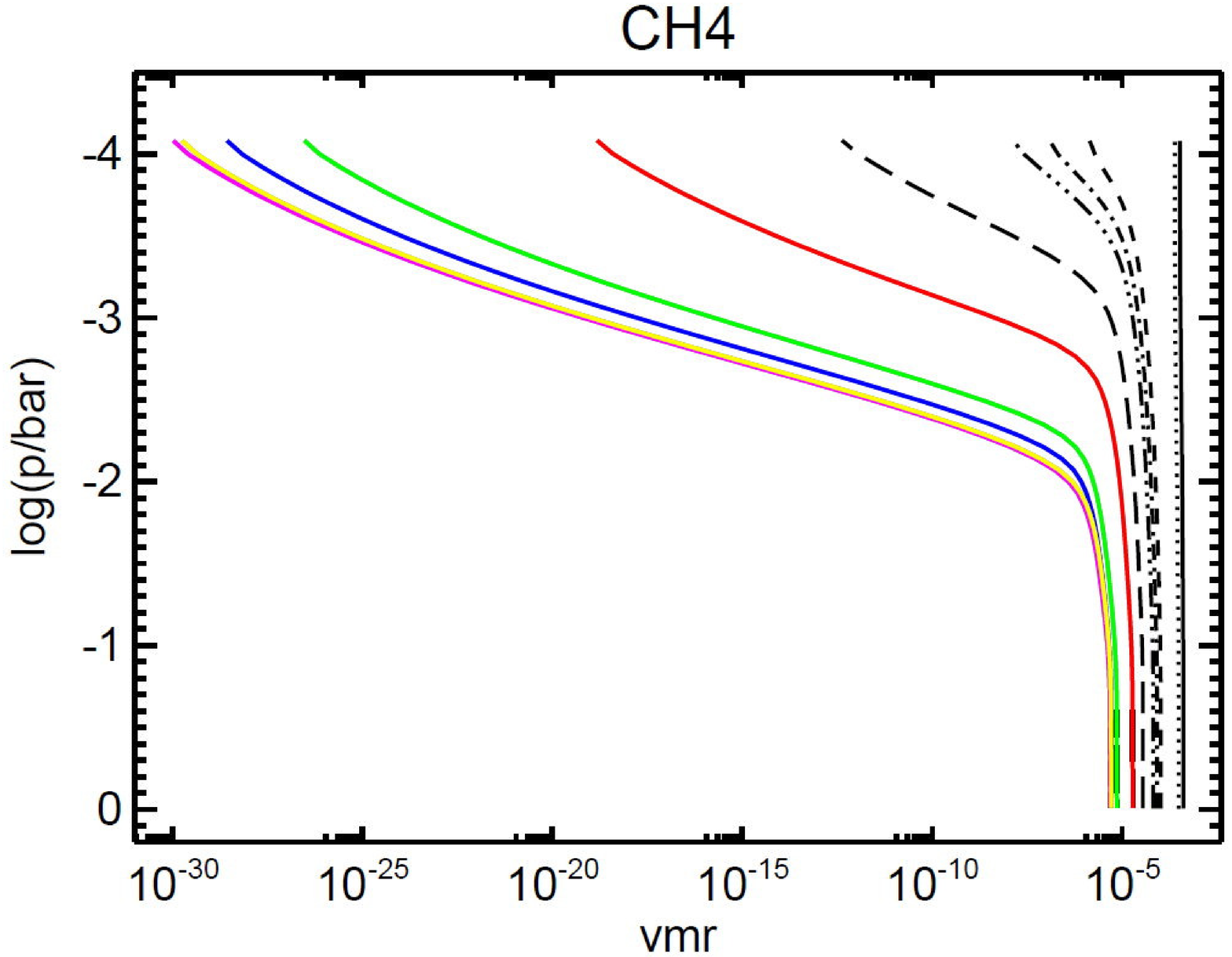} 
  \caption{CH$_4$ profiles of Earth-like  atmosphere runs around the M-dwarf AD Leo calculated by the CAB model. Notation as in Fig. \ref{figure_pTvarO2}.}
  \label{CH4varO2fig}
\end{figure}
\begin{figure}
\centering
\includegraphics[width=8.5cm]{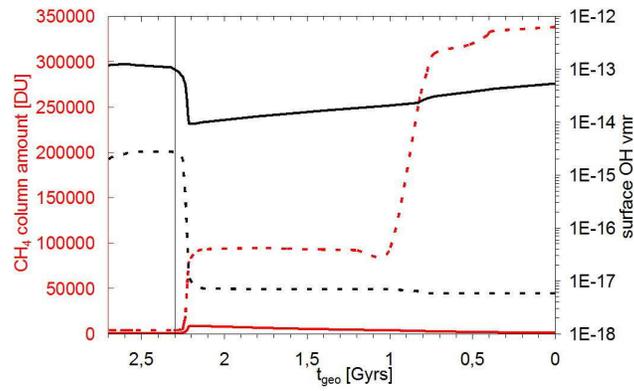} 
  \caption{CH$_4$ column amount [DU] for Earth-like  atmosphere runs around the M-dwarf star AD Leo (dashed red) as a function of geological time $t_{\rm{geo}}$ in comparison to the early-Earth analog planets around the Sun presented in paper I (solid red). The beginning of the GOE at $t_{\rm{geo}} = 2.3$ Gyrs is indicated as a vertical line. Furthermore, surface OH concentrations are given for scenarios around AD Leo in dashed black and around the Sun in solid black.}
  \label{CH4columnfig}
\end{figure}
\begin{figure}
\centering
\includegraphics[width=8.5cm]{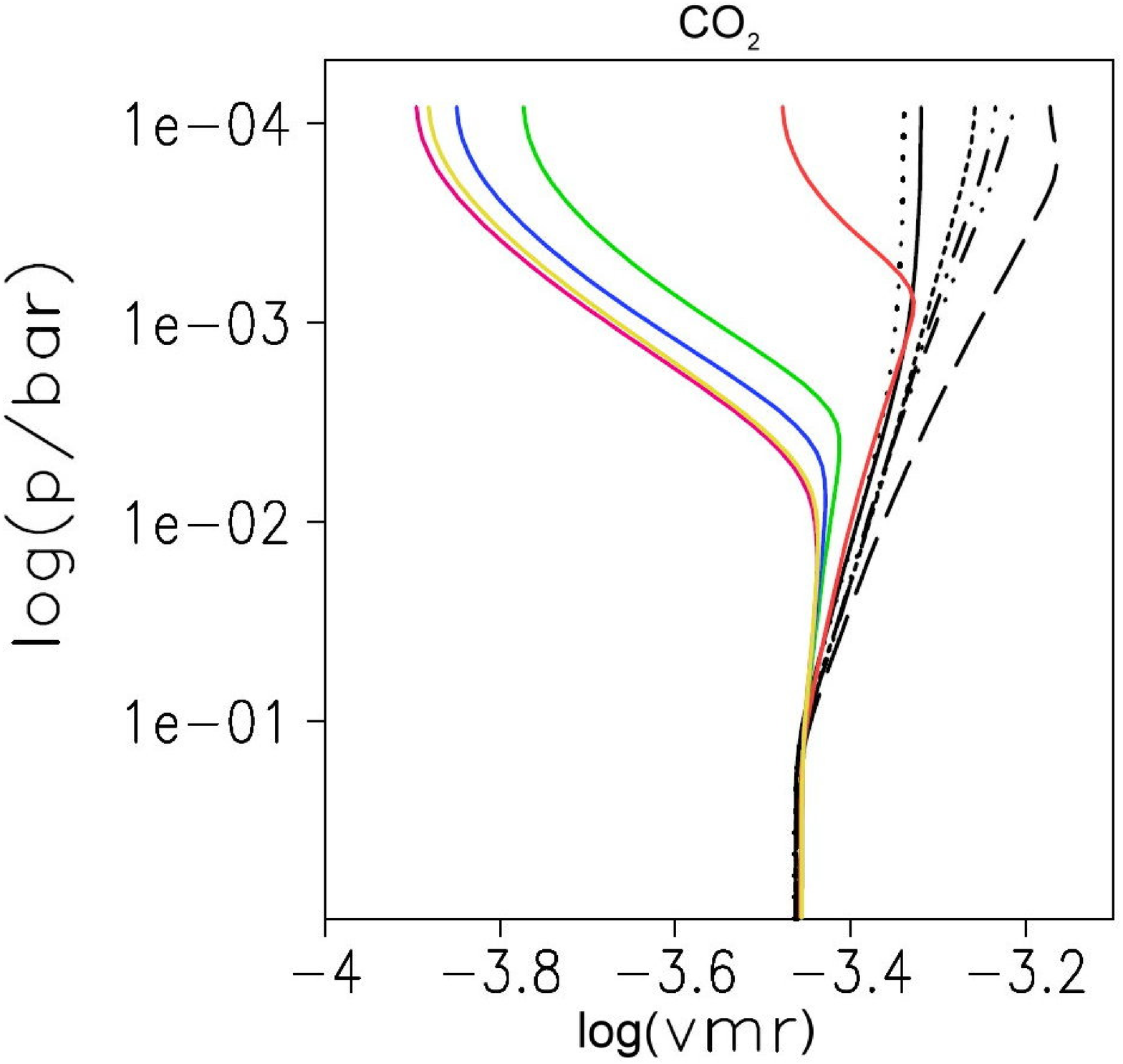} 
  \caption{CO$_2$ profiles of Earth-like  atmosphere runs around the M-dwarf AD Leo calculated by the CAB model. Notation as in Fig. \ref{figure_pTvarO2}.}
  \label{CO2varO2fig}
\end{figure}
\begin{figure}
\centering
\includegraphics[width=8.5cm]{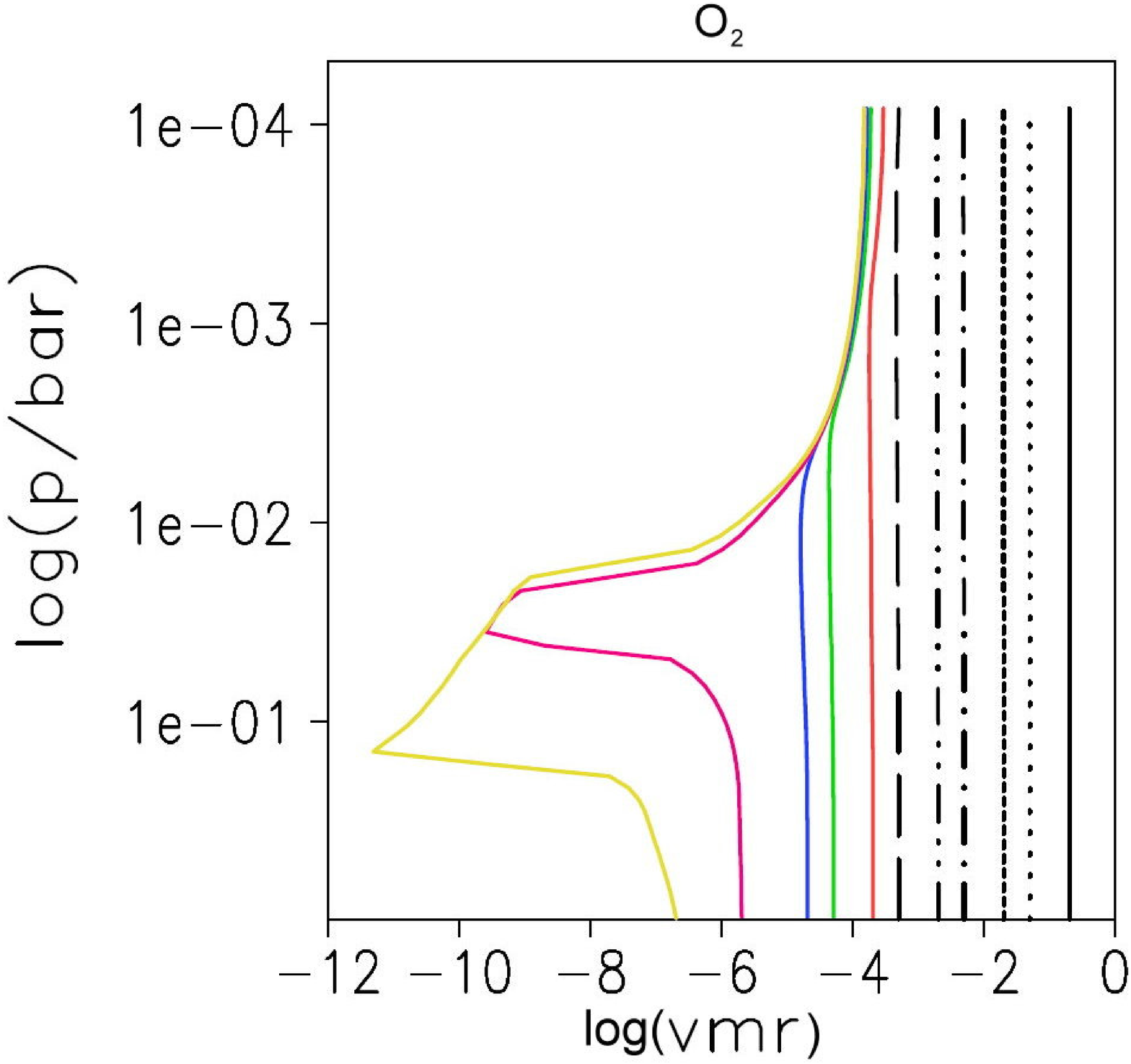} 
  \caption{O$_2$ profiles of Earth-like  atmosphere runs around the M-dwarf AD Leo calculated by the CAB model. Notation as in Fig. \ref{figure_pTvarO2}.}
  \label{O2varO2fig}
\end{figure}

\paragraph{\textbf{Methane - CH$_4$}}
Fig. \ref{CH4varO2fig} shows how the CH$_4$ profile increases on going forwards in time. Increasing the O$_2$ content leads to enhanced O$_3$ formation, hence UV absorption and therefore to less photolytic destruction of CH$_4$ in the upper stratosphere. In comparison to the scenarios presented in paper I, CH$_4$ concentrations are increased for the AD Leo atmosphere scenarios which is directly related to less destruction by OH which is less abundant in the AD Leo atmosphere scenarios (see Fig. \ref{CH4columnfig}) because of decreased production via H$_2$O + O($^1$D) $\rightarrow$ 2OH than for scenarios around the Sun. Although H$_2$O is more abundant in these atmospheres, the mixing ratio of O($^1$D) is reduced in the middle and lower atmosphere because of a decreased production via O$_3$ + h$\nu$ $\rightarrow$ O$_2$ + O($^1$D). Also, the reduced OH concentration leads to less destruction of CO via CO + OH $\rightarrow$ CO$_2$ + H resulting in higher CO concentrations for Earth-like  planetary atmospheres around AD Leo compared to the Sun. The overall column amount of CH$_4$ is also given in Fig. \ref{CH4columnfig}. CH$_4$ increases (as OH decreases) on going forwards in time for the AD Leo atmosphere scenarios. The strong increase of CH$_4$ around the SOE is related to a rather modest decrease in OH and less photolytic destruction.


\paragraph{\textbf{Carbon dioxide - CO$_2$}}
Fig. \ref{CO2varO2fig} shows CO$_2$ profiles of Earth-like  atmosphere scenarios around the M-dwarf star AD Leo. In all atmosphere scenarios CO$_2$ strongly deviates from the iso-profile assumption usually applied in the literature (see e.g. \citealp{segura2003,segura2005,rauer2011}), e.g. at 1 PAL O$_2$ the CO$_2$ concentration increases from the surface to TOA by about 40\%. Below pressures of 0.01 bar, CO$_2$ concentrations increase generally with height due to an increase in UV radiation in the wavelength range favoring OH production. This leads to stronger net production of CO$_2$ via the reaction CO + OH $\rightarrow$ CO$_2$ + H resulting in a decrease in CO towards lower pressures (not shown).
With decreasing O$_2$, CO$_2$ further increases from run to run for ground level O$_2$ concentrations above $2.5\cdot 10^{-3}$ PAL. Below this O$_2$ content (colored lines) the mixing ratio of CO$_2$ decreases towards lower pressures because of enhanced photolytic destruction.

 
\paragraph{\textbf{Molecular oxygen - O$_2$}}
O$_2$ increasingly deviates from the usually applied iso-profile assumption
below $10^{-3}$ PAL O$_2$ as seen in Fig. \ref{O2varO2fig}). The strong decrease in mixing ratio between 0.1 and 0.01 bar for the low O$_2$ atmospheres is correlated to enhanced H abundances which can lead to fast removal of O$_2$ via: H + O$_2$ + M $\rightarrow$ HO$_2$ + M in this pressure range. This also directly impacts the behavior of O$_3$ which is also reduced in this range. In the upper stratosphere O$_2$ production is initiated via the photolysis of CO$_2$ (see for detailed chemical analysis in Section \ref{papanalysis}). 

The altitude dependence of the net ($P - L$) chemical change of O$_2$ from chemical production and destruction is depicted in Fig. \ref{PLO2fig}. In comparison to the atmospheric scenarios around the Sun presented in paper I, O$_2$ is now more strongly destroyed in the upper stratosphere (rather than the troposphere as in the case for the Sun). The net destruction peak which forms for the AD Leo atmosphere scenario with $10^{-3}$ PAL O$_2$ at about atmospheric layer 50 (corresponding to $7.9 \cdot 10^{-4}$ bar) moves to lower altitudes on reducing the surface O$_2$ level concentration. This is directly correlated to a stratospheric maximum in HO$_x$ as a sink for O$_2$ which also moves to lower altitudes (not shown). On reducing the surface O$_2$ concentration the net destruction in the troposphere becomes relatively stronger but is still overall weaker than for the atmosphere scenarios around the Sun presented in paper I. The OH radical which is crucial in the O$_2$ destruction pathways found by PAP for an atmosphere with a surface O$_2$ mixing ratio of $10^{-6}$ PAL O$_2$ around the Sun presented in paper I is strongly reduced in the troposphere for the corresponding AD Leo case with $10^{-6}$ PAL O$_2$ around AD Leo, although other related chemical species such as H, HO$_2$ and H$_2$O$_2$ (see Fig. \ref{HOx_Sun_ADL_fig}) are enhanced. This partitioning occurs due to less photolytic destruction of reservoir species producing OH in the lower atmosphere. In addition to the stronger CO$_2$ photolysis in the upper stratosphere for an Earth-like  planetary atmosphere with $10^{-6}$ PAL O$_2$ around AD Leo, the abundances of the HO$_{\mbox{x}}$ radicals in the upper stratosphere are also enhanced, therefore, increasing the efficiencies of the O$_2$ production pathways which have been presented in paper I.
In general, the resulting atmospheric surface flux of O$_2$ from the column-integrated production and destruction of O$_2$ in the atmosphere which is needed to maintain a specific O$_2$ surface mixing ratio is lower by several hundreds of Tg/yr for the AD Leo atmosphere scenarios than for the Sun cases presented in paper I (the relative change between both surface fluxes varies between 6 to 39\%). This suggests that Earth-like atmospheres around the M-dwarf star AD Leo could be more readily oxidized for a given photosynthetic biosphere.

A detailed chemical analysis of the production and destruction pathways of O$_2$ in the atmosphere of an Earth-like planet orbiting the M-dwarf star AD Leo with 10$^{-6}$ PAL O$_2$ now follows in Section \ref{papanalysis}. Here, we quantify the most dominant production and destruction pathways of O$_2$ by applying PAP.



\subsubsection{Pathway Analysis with respect to O$_2$}\label{papanalysis}

PAP (for details see Section \ref{papsection}) is applied to an exoplanetary atmosphere for the AD Leo scenario with a surface O$_2$ vmr of $10^{-6}$ PAL O$_2$ ($t_{\rm{geo}}=2.688$ Gyrs). Our study represents the first application of PAP in the context of the Earth-like  planet with low O$_2$. As in paper I, we consider only pathways with an individual contribution larger than 1.5\% to the total column production (or destruction) of O$_2$. CO was excluded as a branching point species in order to ensure a consistent treatment of all O$_2$ production and destruction pathways at every atmospheric height. Hence, CO is regarded as a potential source and sink for CO$_2$ in every atmospheric layer.

\paragraph{\textbf{Production pathways and their altitude dependence}}
A summary of the major column-integrated chemical O$_2$ production pathways P1-P7 found by PAP and their percentage contribution to the total colum-integrated O$_2$ production rate is given in Tab. \ref{proPAPpathstab} for an Earth-like  atmosphere with $10^{-6}$ PAL O$_2$ orbiting in the HZ of the M-dwarf star AD Leo in comparison to the pathways and their percentage contribution found in paper I for early Earth analog planets around the Sun.
As in paper I, these pathways can be broadly categorized into three classes, namely:
\begin{itemize}
	\item class PA: O$_2$ production via net reaction 2CO$_2$ $\rightarrow$ O$_2$ + 2CO (initiated by CO$_2$ photolysis)
		\begin{itemize}
			\item subclass PA1: catalyzed by HO$_x$
				\begin{itemize}
					\item PA1a (P1): catalyzed by O$_3$
					\item PA1b (P2): without O$_3$
				\end{itemize}
			\item subclass PA2 (P3): without HO$_x$
\end{itemize}		 
	\item class PB: O$_2$ production via net reaction 2O $\rightarrow$ O$_2$
		\begin{itemize}
			\item subclass PB1: catalyzed by HO$_x$
				\begin{itemize}
					\item PB1a (P4): catalyzed by O$_3$ (new for AD Leo)
					\item PB1b (P7): without O$_3$
				\end{itemize}
			\item subclass PB2 (P6): without HO$_x$
\end{itemize}		
	\item class PC (P5): O$_2$ production via net reaction H$_2$O + CO$_2$ $\rightarrow$ O$_2$ + 2H + CO (new for AD Leo)
\end{itemize}
Note that pathways of subclasses PB1a and PA1a, PB1b and PA1b and PB2 and PA2 are closely related. They differ in providing O by either in-situ photolysis of CO$_2$ or delivery of O by diffusion from photolyzed CO$_2$ in other atmospheric layers.
 
\begin{figure}
\centering
\includegraphics[width=8.5cm]{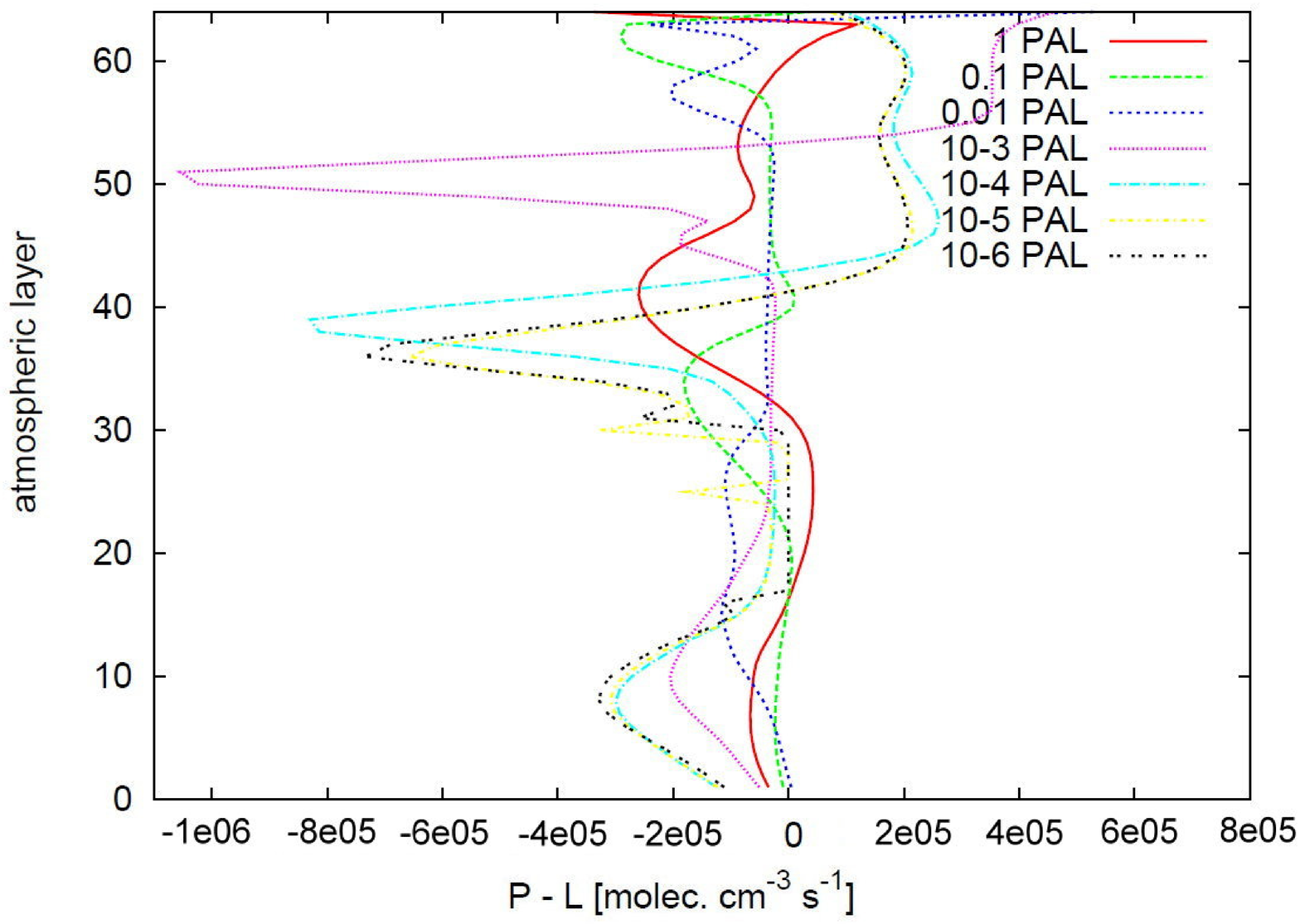} 
  \caption{Atmospheric net $(P-L)$ chemical change of O$_2$ calculated by the CAB model for Earth-like  atmosphere scenarios around the M-dwarf AD Leo. Notation: solid red: 1 PAL O$_2$ (modern Earth), dashed green: $10^{-1}$ PAL ($t_{\rm{geo}}=2.18$ Gyrs), dotted dark blue: $10^{-2}$ PAL ($t_{\rm{geo}}=2.22$ Gyrs), dotted purple: $10^{-3}$ PAL ($t_{\rm{geo}}=2.24$ Gyrs), dot-dashed cyan: $10^{-4}$ PAL ($t_{\rm{geo}}=2.28$ Gyrs), dot-dashed yellow: $10^{-5}$ PAL ($t_{\rm{geo}}=2.59$ Gyrs), dot-dot black: $10^{-6}$ PAL ($t_{\rm{geo}}=2.69$ Gyrs). Note that values are plotted for each atmospheric layer instead of pressure because every scenario presented refers to a different pressure grid due to pressure adjustment calculated from the evaporation of H$_2$O at the surface.}
  \label{PLO2fig}
\end{figure}

\begin{figure}
\hfill\\
\centering
\includegraphics[width=8.5cm]{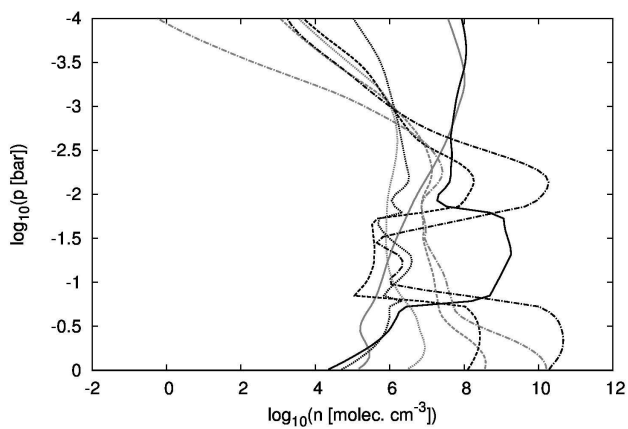}
  \caption{Profiles of H (black solid), OH (black dotted), HO$_2$ (black dashed) and H$_2$O$_2$ (black dot-dashed) for an Earth-like  atmosphere scenario around the M-dwarf AD Leo with $10^{-6}$ PAL O$_2$. Profiles in grey show the corresponding case for the Sun for comparison.}\label{HOx_Sun_ADL_fig}
  \end{figure}

\begin{figure}
\centering
\includegraphics[width=8.5cm]{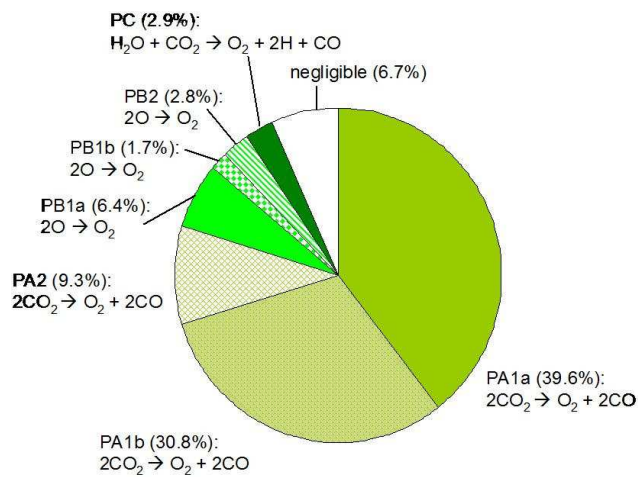}
  \caption{O$_2$ production classes (see Tab. \ref{proPAPpathstab}) and their contributions to the total column-integrated O$_2$ production rate via pathways calculated by PAP for an Earth-like atmosphere with a surface O$_2$ vmr of $10^{-6}$ PAL O$_2$.}
  \label{piePfig}
\end{figure}

Fig. \ref{piePfig} shows the relative contributions of the pathways mentioned above to the total column-integrated production rate of O$_2$. Pathways P1, P2, P3, P6 and P7 are also present in the paper I analysis but their relative contributions are very different. In Earth-like  atmospheres around the M-dwarf AD Leo the largest contribution originates from subclass PA1. The contribution of pathway P1 (P7 in paper I, i.e. I-P7, the same notation is used for all other pathways found in paper I) is now increased from 2.3\% to 39.6\% because of enhanced O$_3$ and HO$_x$ in comparison to paper I (for O$_3$ see their Fig. 7 and for HO$_x$ see Fig. \ref{HOx_Sun_ADL_fig}). Pathway P2 is slightly increased whereas for P3 (I-P3 in paper I) a distinct relative decrease from 20.7\% to 9.3\% is observed which is related to the fact that the total O$_2$ production rate is 3 times higher in the case of Earth-like  planets orbiting AD Leo (compare Fig. 17 in paper I with Fig. \ref{PLO2fig}). There are two new pathways present, namely P4 and P5. P4 contains similar reactions as in P1 but is not initiated by the photolysis of CO$_2$ and subsequent relaxation of O($^1$D) only depending on the presence of O delivered by diffusion.
Pathway P5 contains the photolysis of CO$_2$ and H$_2$O releasing O and OH which then recombine to form O$_2$ and H.
The percetange contributions of pathways P6 and P7 are similar to the results of paper I.    

\begin{figure}
\centering
\includegraphics[width=8.5cm]{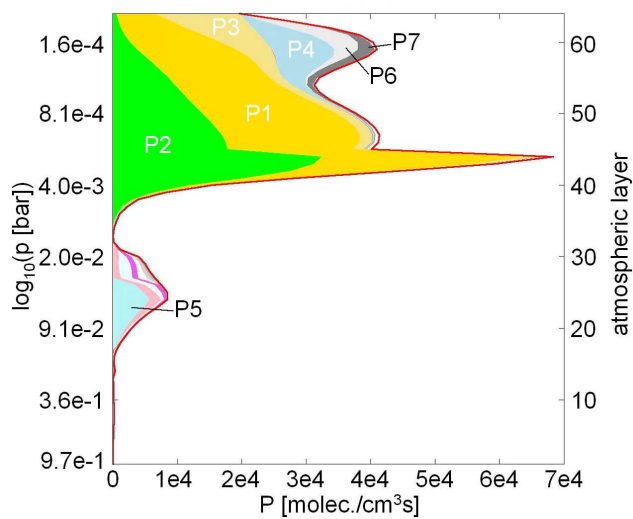} 
  \caption{Altitude dependence of O$_2$ production pathways P1 to P7 (see Tab. \ref{proPAPpathstab}) for an Earth-like atmosphere with a ground level O$_2$ concentration of 10$^{-6}$ PAL around the M-dwarf AD Leo.}
  \label{allpro1emin6fig}
\end{figure}

Fig. \ref{allpro1emin6fig} shows the altitude dependence of the O$_2$ production pathways for an Earth-like  planet with $10^{-6}$ PAL O$_2$ around AD Leo. The total production rate of O$_2$ (solid red line) in the CAB model is also plotted. The sum of all pathways found by PAP amounts to 98.3\% of the total O$_2$ production rate.
Fig. \ref{allpro1emin6fig} indicates that the strongest production of O$_2$ occurs in the upper atmosphere mainly from pathways initiated by CO$_2$ photolysis where UV radiation is strong. A small contribution to the total production of O$_2$ arises from pathway P5 which additionally photolyzes H$_2$O due to strong shortwave radiation penetrating deeply into the atmosphere.

\paragraph{\textbf{Destruction pathways and their altitude dependence}}
A summary of the major column-integrated chemical O$_2$ destruction pathways L1-L9 found by PAP and their percentage contribution to the total column-integrated O$_2$ destruction rate is given in Tab. \ref{lossPAPpathstab} for an Earth-like  atmosphere with $10^{-6}$ PAL O$_2$ orbiting around the M-dwarf star AD Leo in comparison to the pathways and their percentage contribution found in paper I for early Earth analog planets around the Sun.
As in paper I, these pathways can be categorized into 4 classes, namely:
\begin{itemize}
	\item class LA: O$_2$ destruction via net reaction O$_2$ + 2CO $\rightarrow$ 2CO$_2$ (CO oxidation)
		\begin{itemize}
			\item subclass LA1 (L1): catalyzed by HO$_x$ but without photolysis
			\item subclass LA2: catalyzed by HO$_x$ but with photolysis of 
			\begin{itemize}
					\item LA2a (L2): H$_2$O$_2$
					\item LA2b (L4, L6, L7): O$_2$
				\end{itemize}
\end{itemize}		 
	\item class LB: O$_2$ destruction via CH$_4$ oxidation pathways
		\begin{itemize}
			\item subclass LB1 (L3): net reaction 3O$_2$ + 2CH$_4$ $\rightarrow$ 4H$_2$O + 2CO
			\item subclass LB2 (L8): net reaction O$_2$ + CH$_4$ $\rightarrow$ CH$_3$OOH (new for AD Leo)
\end{itemize}		
	\item class LC (L9): O$_2$ destruction via net reaction O$_2$ + 2H$_2$ $\rightarrow$ 2H$_2$O
	\item class LD (L5): O$_2$ destruction via net reaction O$_2$ + H$_2$O + CO $\rightarrow$ H$_2$O$_2$ + CO$_2$
\end{itemize}

\begin{figure}
\centering
\includegraphics[width=8.5cm]{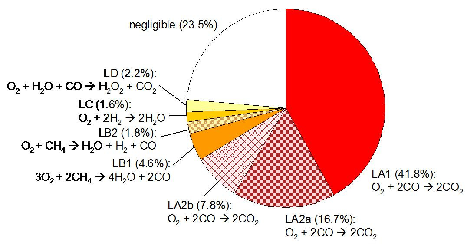} 
  \caption{As for Fig. \ref{piePfig} but for destruction (see also Tab. \ref{lossPAPpathstab}).}
  \label{pieLfig}
\end{figure}

Fig. \ref{pieLfig} shows the relative contributions of the pathways mentioned above to the total column-integrated destruction rate of O$_2$. Pathways L1, L2, L3, L4, L5, L6 and L9 are also present in early-Earth analog atmosphere around the Sun presented in paper I but their relative contributions are very different.
In an Earth-like  atmosphere with $10^{-6}$ PAL O$_2$ around AD Leo the largest contribution (66.3\%) arises from class LA, i.e. oxidation of CO. Within this class, pathway L1 dominates the column-integrated O$_2$ destruction with 41.8\% because of enhanced HO$_x$ species released by strong photolysis of H$_2$O in this pressure range whereas this pathway, L8 in paper I, i.e. I-L8 (the same notation is used for all other pathways found in paper I), is rather minor for the scenario presented there.
The contribution of pathway L2 (I-L1) is now decreased from 28.5\% to 16.7\% because of less OH in the lower atmosphere compared to paper I (see Fig. \ref{HOx_Sun_ADL_fig}). The contributions of other pathways from e.g. CH$_4$ and H$_2$O oxidation are rather small.  
There are two new pathways present, namely L7 and L8. 
L7 is additionally catalyzed by O$_3$ which is more abundant in this altitude range than in the early Earth analog planet with $10^{-6}$ PAL O$_2$ around the Sun presented in paper I.
Pathway L8 is initialized by the photolysis of H$_2$O resulting in the oxidation of CH$_4$ by forming CH$_3$OOH.
 
\begin{figure}
\centering
\includegraphics[width=8.5cm]{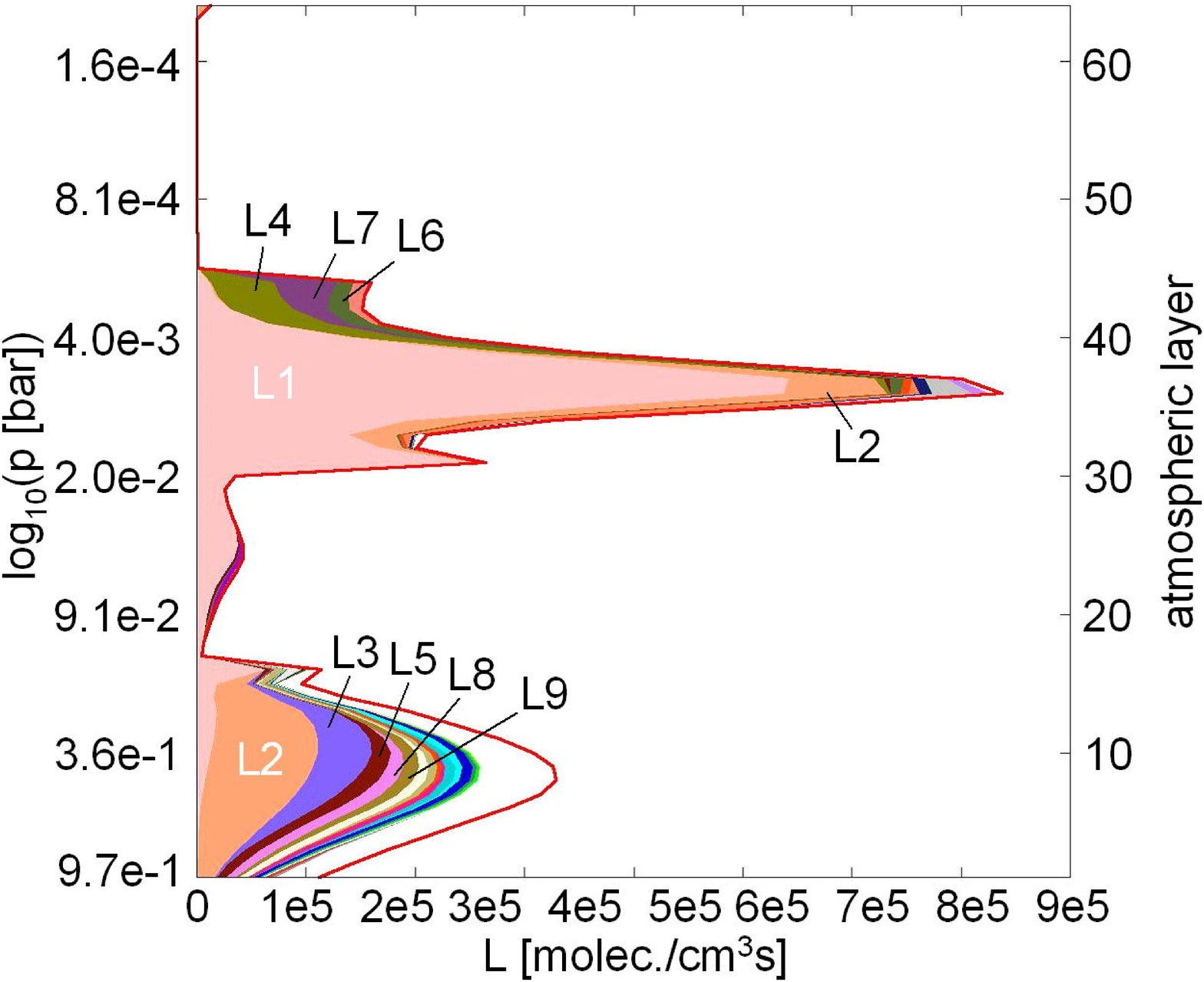} 
  \caption{As for Fig. \ref{allpro1emin6fig} but for destruction (see also Tab. \ref{lossPAPpathstab}).}
  \label{alldes1emin6fig}
\end{figure}

The altitude dependance of the presented destruction pathways for an Earth-like  atmosphere with $10^{-6}$ PAL O$_2$ around the M-dwarf AD Leo is given in Fig. \ref{alldes1emin6fig}. The total destruction rate of O$_2$ is also shown (solid red line). The sum of all pathways found by PAP amounts to 89.9\% of the total O$_2$ destruction rate. The strongest destruction of O$_2$ occurs in middle atmosphere followed by a smaller contribution in the troposphere where destruction is composed of a large number of pathways individually contributing less than 1.5\%. 
The gap between both regimes can be attributed a minimum in CO (not shown) as well as HO$_2$ and H$_2$O$_2$ (see Fig. \ref{HOx_Sun_ADL_fig}) in this pressure region.

The minimum in O$_2$ vmr which is seen in Fig. \ref{O2varO2fig} results from low O$_2$ production and destruction rates (although there is a local maximum in O$_2$ production). From Fig. \ref{PLO2fig} one can deduce that these rates cancel each other out in the middle atmosphere (dot-dot black line), hence we can conclude that the increase in O$_2$ vmr (see Fig. \ref{O2varO2fig}) with altitude is due to diffusion from above atmospheric layers where O$_2$ is produced abiotically from CO$_2$ photolysis. 

In summary, for an Earth-like  atmosphere with $10^{-6}$ PAL O$_2$ around the M-dwarf AD Leo, O$_2$ can be produced in-situ from CO$_2$ photolysis followed by HO$_x$ catalyzed recombination of O to form O$_2$ in the upper atmosphere. This result was already proposed by other studies such as e.g. \citet{selsis2002,hu2012,doma2014} but has now been quantified by our work. 
The destruction of O$_2$ is composed of different oxidation pathways, mainly of CO as already shown in paper I. However, in the present paper the main contribution arises from the middle atmosphere where the H$_2$O photolysis is the strongest followed by the tropospheric pathways destroying O$_2$. Note too that as in paper I the biological O$_2$ flux originating from the surface is orders of magnitude higher than the abiotic in-situ production in the atmosphere.

\subsection{Biogeochemical Modeling}\label{biosphere}
\begin{figure}
	\centering
\includegraphics[width=8.5cm]{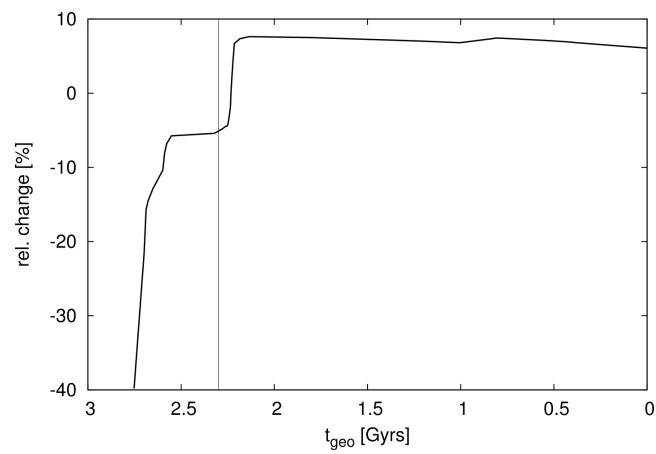} 
  \caption{Calculated relative change in \% of the input from oxygenic photosynthesis, $N$, in Tg O$_2$/yr for Earth-like  atmosphere scenarios around AD Leo compared to the Sun as a function of geological time $t_{\rm{geo}}$ .}
  \label{N_rconstfig}
	\end{figure}

The CAB model calculates how much input from oxygenic photosynthesis, $N$ (NPP), is needed to maintain the surface flux $\Phi^{atm}_{\rm{O}_2}$ calculated by the atmospheric chemistry module for a given surface vmr of O$_2$.
For M-dwarf star planets PAR is reduced resulting in less net
primary productivity by oxygenic photosynthesis (see comments above and Kiang et al. 2007).
The CAB model contains no information on the sensitivity of photosynthesis\footnote{Photosynthetic (light harvesting) pigments peak in absorbance at wavelengths where light intensity on the planetary surface is highest (Kiang et al. 2007).} to the stellar spectrum. This issue however is beyond the scope of our work.
Fig. \ref{N_rconstfig} shows the relative change in \% in input from oxygenic photosynthesis, $N$, over geological time for constant input of Fe$^{2+}$ for anoxygenic photosynthesis, $r$, if one assumes Earth's biosphere evolution for an Earth-like  planet orbiting AD Leo. 

In regime I, before about $t_{\rm{geo}}=2.24$ Gyrs ago (below roughly $10^{-3}$ PAL O$_2$) the input from oxygenic photosynthesis, $N$, which is needed to maintain a specified O$_2$ surface mixing ratio, is less than for planets around the Sun presented in paper I. The lower the O$_2$ content (going further back in time) the lower the $N$ compared to the Sun. In this regime $N$ is sensitive to the O$_2$ surface flux, hence the chemical nature of the atmosphere. The reduced $N$ implies that even for reduced NPP, the accumulation of atmospheric O$_2$ by photosynthesizers might still be favorable for Earth-like  planets around AD Leo because of the less destructive nature of the atmosphere for O$_2$ (see Fig. \ref{PLO2fig} for net ($P - L$) chemical change of O$_2$ for Earth-like  atmospheres around star AD Leo) despite higher concentrations of reduced gases such as e.g. H$_2$, CH$_4$ and CO. This is driven by weaker tropospheric OH, hence weaker loss of O$_2$ via CO oxidation and stronger abiotic O$_2$ production due to CO$_2$ photolysis in the AD Leo case.

In regime II, after about $t_{\rm{geo}}=2.24$ Gyrs ago (above around $10^{-3}$ PAL O$_2$), $N$ is more sensitive to the O$_2$ partial pressure and surface temperature rather than atmospheric chemistry, hence O$_2$ surface flux (as was shown in paper I). Although O$_2$ surface fluxes are lower than compared to the Sun, $N$ is increased by about 7\% due to higher $p_{\mbox{surf}}$ and $T_{\mbox{surf}}$ which is correlated to the greenhouse effect of gases such as H$_2$O and CH$_4$.


\section{Conclusion}\label{resultsection}
This is the first study with our newly developed CAB model \citep{gebauer2017} addressing Earth-like  planets with varying atmospheric O$_2$ content orbiting in the HZ around the M-dwarf star AD Leo.

Despite higher atmospheric CH$_4$, CO and H$_2$ concentrations than for scenarios presented in paper I for the Sun, O$_2$ is overall less destroyed in the AD Leo case e.g. because of smaller OH concentrations in the troposphere in comparison to the Sun scenarios. Hence this leads to a weakening in the OH catalyzed loss of O$_2$ via CO oxidation.

For the AD Leo case with 1 PAL O$_2$ the CO$_2$ concentrations increase towards higher altitudes by 40\% implying that the usually applied iso-profile assumption (see e.g. \citealp{segura2003,rauer2011}) is not valid in such atmospheres.

Lowering the ground level O$_2$ concentration led to in-situ production of O$_2$ from CO$_2$ photolysis in the upper stratosphere resulting in an increase in O$_2$ abundances and strong depletion of CO$_2$ mixing ratios towards higher altitudes in these atmospheres. A minimum in O$_2$ concentration becomes visible for the low O$_2$ atmosphere scenarios which is related to the different partitioning of HO$_x$ species over atmospheric height. 
Generally, the atmospheres of Earth-like  planets around the M-dwarf star AD Leo with different surface O$_2$ concentrations are less destructive towards O$_2$ than for the scenarios presented in paper I for the Sun. Results therefore suggest
that the accumulation of O$_2$ in Earth-like  planets around AD Leo might be chemically more favorable compared to Earth-like planets orbiting around G-type stars (such as the Sun). 

The current work presents the first application of PAP quantifying the production and destruction pathways of O$_2$ for an Earth-like  planet with $10^{-6}$ PAL O$_2$ orbiting the M-dwarf star AD Leo.
It was shown the the main production arises from CO$_2$ photolysis followed by catalytic HO$_x$ reactions in the upper atmosphere whereas the strongest destruction arises not from the troposphere, as was the case in paper I, but instead in the middle atmosphere where H$_2$O photolysis is stronger.

If one assumes the same biosphere evolution for Earth-like  planets around the M-dwarf star AD Leo as on Earth then the net primary productivity, $N$, from oxygenic photosynthesis is strongly decreased for Earth-like 
planets with a surface O$_2$ mixing ratio below about $10^{-3}$ PAL O$_2$. This implies that the productivity required for a given atmospheric O$_2$ abundance is less strong than in paper I. For atmospheres having ground
level O$_2$ concentrations above $10^{-3}$ PAL O$_2$ $N$ is slightly increased in comparison to the Sun
scenarios due to the higher surface temperatures, hence partial pressures of O$_2$.

In terms of next generation missions aiming at the characterization of Earth-like  planets the results of this work imply that the feasability of the presence of atmospheric O$_2$, hence O$_3$, might be enhanced because a potential "Great Oxidation Event" might have had occured earlier in geological time since the oxidative capacity driven by the stellar input spectrum is decreased. However, the distinguishabilty of biotic and abiotic O$_2$ and O$_3$ needs further investigation.

\newpage

\noindent \textbf{Acknowledgments:}\\
Stefanie Gebauer acknowledges support by the DFG SPP 1833 "Building a Habitable Earth".

This research was supported by the \emph{Helmholtz Gemeinschaft}
  through the research alliance "Planetary Evolution and Life".


\noindent \textbf{Author Disclosure Statement:}\\
No competing financial interests exist.

\bibliographystyle{agufull08}
\bibliography{Bibo_1D_2}

\newpage

\begin{longtable} {c c c c c}

\caption{Summary of dominant chemical production (P) pathways of O$_2$ found by PAP and their percentage contribution to the total column-integrated O$_2$ production rate for an Earth-like  atmosphere with a surface O$_2$ vmr of $10^{-6}$ PAL O$_2$. Only pathways that contribute $>1.5$\% of the total column-integrated production rate are shown. For an explanation of classes and subclasses see text. Values in parentheses indicate the pathway number, i.e. I-Px, and percentage contribution from paper I.}\\
\hline
class & subclass & pathway number & contribution $[\%]$ & pathway\\
\hline 
\endfirsthead
\hline	
class & subclass & pathway number & contribution $[\%]$ & pathway\\
\hline 
\endhead
\hline
\endfoot
\hline
\endlastfoot


 PA  & PA1a & P1 (I-P7) & 39.6 (2.3) & 2$\cdot$(CO$_2$ + h$\nu$ $\rightarrow$ CO + O($^1$D))\\			
  & &   & & 2$\cdot$(O($^1$D) + N$_2$ $\rightarrow$ O + N$_2$)\\
  & &	& & O + O$_2$ + M $\rightarrow$ O$_3$ + M\\			
  & &	& & OH + O $\rightarrow$ H + O$_2$\\			
  & &	& & H + O$_3$ $\rightarrow$ OH + O$_2$\\			
  & &	& & \textbf{net: 2CO$_2$ $\rightarrow$ O$_2$ + 2CO}\\\hline
 
 & PA1b & P2 (I-P1) & 30.8 (28.4) & 2$\cdot$(CO$_2$ + h$\nu$ $\rightarrow$ CO + O($^1$D))\\			
  & &	& & 2$\cdot$(O($^1$D) + N$_2$ $\rightarrow$ O + N$_2$)\\			
  & &	& & HO$_2$ + O $\rightarrow$ OH + O$_2$\\			
  & &	& & OH + O $\rightarrow$ H + O$_2$\\			
  & &	& & H + O$_2$ + M $\rightarrow$ HO$_2$ + M\\			
  & &	& & \textbf{net: 2CO$_2$ $\rightarrow$ O$_2$ + 2CO}\\\hline
 
  	
 & PA2 & P3 (I-P3) & 9.3 (20.7) & 2$\cdot$(CO$_2$ + h$\nu$ $\rightarrow$ CO + O($^1$D))\\		
  & &	& & 2$\cdot$(O($^1$D) + N$_2$ $\rightarrow$ O + N$_2$)\\			
  & &	& & O + O + M $\rightarrow$ O$_2$ + M\\			
  & &	& & \textbf{net: 2CO$_2$ $\rightarrow$ O$_2$ + 2CO}\\\hline
 
  	 	  
 PB & PB1a & P4 (-) & 6.4 (-) & O + O$_2$ + M $\rightarrow$ O$_3$ + M\\				
  & &	& & H + O$_3$ $\rightarrow$ OH + O$_2$\\					
  & &	& & OH + O $\rightarrow$ H + O$_2$\\		
  & &	& & \textbf{net: 2O $\rightarrow$ O$_2$}\\\hline  	 	  
  	 	  	
  & PB1b & P7 (I-P6) & 1.7 (3.9) & OH + O $\rightarrow$ H + O$_2$\\					
  & &	& & H + O$_2$ + M $\rightarrow$ HO$_2$ + M\\					
  & &	& & HO$_2$ + O $\rightarrow$ OH + O$_2$\\		
  & &	& & \textbf{net: 2O $\rightarrow$ O$_2$}\\\hline
  
 & PB2 & P6 (I-P5) & 2.8 (5.3) & O + O + M $\rightarrow$ O$_2$ + M\\			
  & &	& & \textbf{net: 2O $\rightarrow$ O$_2$}\\\hline

 PC &  & P5 (-) & 2.9 (-) & CO$_2$ + h$\nu$ $\rightarrow$ CO + O\\				
  & &	& & H$_2$O + h$\nu$ $\rightarrow$ H + OH\\					
  & &	& & OH + O $\rightarrow$ H + O$_2$\\		
  & &	& & \textbf{net: H$_2$O + CO$_2$ $\rightarrow$ O$_2$ + 2H + CO}
\label{proPAPpathstab}
\end{longtable}

\newpage 

\begin{longtable} {c c c c c}

\caption{As for Tab. \ref{proPAPpathstab} but for O$_2$ destruction (L) pathways.}\\
\hline
class & subclass & pathway number & contribution $[\%]$ & pathway\\
\hline 
\endfirsthead
\hline	
class & subclass & pathway number & contribution $[\%]$ & pathway\\
\hline 
\endhead
\hline
\endfoot
\hline
\endlastfoot


 LA & LA1 & L1 (I-L8) & 41.8 (3.4) & H + O$_2$ + M $\rightarrow$ HO$_2$ + M\\
   & & & & H + HO$_2$ $\rightarrow$ OH + OH\\			
   & & & & 2$\cdot$(CO + OH $\rightarrow$ CO$_2$ + H)\\					
   & & & & \textbf{net: O$_2$ + 2CO $\rightarrow$ 2CO$_2$}\\\hline
 
   & LA2a & L2 (I-L1) & 16.7 (28.5) & 2$\cdot$(H + O$_2$ + M $\rightarrow$ HO$_2$ + M)\\		
   & & & & HO$_2$ + HO$_2$ $\rightarrow$ H$_2$O$_2$ + O$_2$\\			
   & & & & H$_2$O$_2$ + h$\nu$ $\rightarrow$ OH + OH\\			
   & & & & 2$\cdot$(CO + OH $\rightarrow$ CO$_2$ + H)\\			
   & & & & \textbf{net: O$_2$ + 2CO $\rightarrow$ 2CO$_2$}\\\hline

   & LA2b & L4 (I-L5) & 3.9 (5.0) & O$_2$ + h$\nu$ $\rightarrow$ O + O($^1$D)\\
   & & & & O($^1$D) + N$_2$ $\rightarrow$ O + N$_2$\\			
   & & & & 2$\cdot$(HO$_2$ + O $\rightarrow$ OH + O$_2$)\\			
   & & & & 2$\cdot$(CO + OH $\rightarrow$ CO$_2$ + H)\\			
   & & & & 2$\cdot$(H + O$_2$ + M $\rightarrow$ HO$_2$ + M)\\			
   & & & & \textbf{net: O$_2$ + 2CO $\rightarrow$ 2CO$_2$}\\\hline

   & LA2b & L6 (I-L3) & 2.0 (7.4) & O$_2$ + h$\nu$ $\rightarrow$ O + O\\			
   & & & & 2$\cdot$(HO$_2$ + O $\rightarrow$ OH + O$_2$)\\			
   & & & & 2$\cdot$(CO + OH $\rightarrow$ CO$_2$ + H)\\			
   & & & & 2$\cdot$(H + O$_2$ + M $\rightarrow$ HO$_2$ + M)\\			
   & & & & \textbf{net: O$_2$ + 2CO $\rightarrow$ 2CO$_2$}\\\hline

   & LA2b & L7 (-) & 1.9 (-) & O$_2$ + h$\nu$ $\rightarrow$ O + O($^1$D)\\			
   & & & & O($^1$D) + N$_2$ $\rightarrow$ O + N$_2$\\
   & & & & 2$\cdot$(O + O$_2$ + M $\rightarrow$ O$_3$ + M)\\			
   & & & & 2$\cdot$(H + O$_3$ $\rightarrow$ OH + O$_2$)\\			
   & & & & 2$\cdot$(CO + OH $\rightarrow$ CO$_2$ + H)\\			
   & & & & \textbf{net: O$_2$ + 2CO $\rightarrow$ 2CO$_2$}\\\hline

  
    	
LB & LB1 & L3 (I-L4) & 4.6 (5.4) & H$_2$O$_2$ + h$\nu$ $\rightarrow$ OH + OH\\		
  & & & & 2$\cdot$(CH$_4$ + OH $\rightarrow$ CH$_3$ + H$_2$O)\\	
  & & & & 2$\cdot$(CH$_3$ + O$_2$ + M $\rightarrow$ CH$_3$O$_2$ + M)\\	
  & & & & 2$\cdot$(CH$_3$O$_2$ + HO$_2$ $\rightarrow$ CH$_3$OOH + O$_2$)\\	
  & & & & 2$\cdot$(CH$_3$OOH + h$\nu$ $\rightarrow$ H$_3$CO + OH)\\	
  & & & & 2$\cdot$(H$_2$CO + OH $\rightarrow$ H$_2$O + HCO)\\	
  & & & & 2$\cdot$(HCO + O$_2$ $\rightarrow$ HO$_2$ + CO)\\	
  & & & & HO$_2$ + HO$_2$ $\rightarrow$ H$_2$O$_2$ + O$_2$\\	
  & & & & 2$\cdot$(H$_3$CO + O$_2$ $\rightarrow$ H$_2$CO + HO$_2$)\\	
  & & & & \textbf{net: 3O$_2$ + 2CH$_4$ $\rightarrow$ 4H$_2$O + 2CO}\\\hline
  
  & LB2 & L8 (-) & 1.8 (-) & H$_2$O + h$\nu$ $\rightarrow$ H + OH\\		
  & & & & H + O$_2$ + M $\rightarrow$ HO$_2$ + M\\	
  & & & & CH$_4$ + OH $\rightarrow$ CH$_3$ + H$_2$O\\	
  & & & & CH$_3$ + O$_2$ + M $\rightarrow$ CH$_3$O$_2$ + M\\	
  & & & & CH$_3$O$_2$ + HO$_2$ $\rightarrow$ CH$_3$OOH + O$_2$\\	
  & & & & \textbf{net: O$_2$ + CH$_4$ $\rightarrow$ CH$_3$OOH}\\\hline  


   
LC & & L9 (I-L9) & 1.6 (3.0) & 2$\cdot$(H + O$_2$ + M $\rightarrow$ HO$_2$ + M)\\
   & & & & HO$_2$ + HO$_2$ $\rightarrow$ H$_2$O$_2$ + O$_2$\\
   & & & & H$_2$O$_2$ + h$\nu$ $\rightarrow$	OH + OH\\		
   & & & & 2$\cdot$(H$_2$ + OH $\rightarrow$ H$_2$O + H)\\					
   & & & & \textbf{net: O$_2$ + 2H$_2$ $\rightarrow$ 2H$_2$O}\\\hline 
    
LD & & L5 (I-L11) & 2.2 (1.7) & CO + OH $\rightarrow$ CO$_2$ + H\\
   & & & & H$_2$O + h$\nu$ $\rightarrow$ H + OH\\
   & & & & 2$\cdot$(H + O$_2$ + M $\rightarrow$ HO$_2$ + M)\\		
   & & & & HO$_2$ + HO$_2$ $\rightarrow$ H$_2$O$_2$ + O$_2$\\					
   & & & & \textbf{net: O$_2$ + H$_2$O + CO $\rightarrow$ H$_2$O$_2$ + CO$_2$}
\label{lossPAPpathstab}
\end{longtable}
    
\newpage    
    
\end{document}